\documentclass[11pt]{article}

\usepackage{amsmath,amssymb}
\usepackage{mathrsfs}
\usepackage{graphicx,color}
\usepackage{geometry,setspace}
\usepackage[english]{babel}
\usepackage[T1]{fontenc}
\usepackage[utf8]{inputenc}
\usepackage[labelfont=bf,labelsep=space]{caption}
\usepackage{subfigure}
\usepackage{setspace}
\usepackage{nicefrac}
\usepackage{hyperref}
\usepackage[capitalise]{cleveref}

\definecolor {darkred}{rgb}{0.6,0,0}
\definecolor {orangered}{rgb}{0.85,0.325,0.098}
\definecolor {orange}{rgb}{1,0.5,0}
\definecolor {darkgreen}{rgb}{0,0.6,0.2}
\definecolor {lightgreen}{rgb}{0.5,0.8,0.35}
\definecolor {lightblue}{rgb}{0.686,0.776,0.914}
\definecolor {mediumblue}{rgb}{0,0.5,0.9}
\definecolor {purple}{rgb}{0.5,0,1}
\definecolor {darkpurple}{rgb}{0.251,0,0.502}
\definecolor {colModelD}{rgb}{0,0.549,0.278}
\definecolor {colModelJ}{rgb}{0.85,0.325,0.098}
\definecolor {lightgray}{rgb}{0.5,0.5,0.5}

\newcommand {\divv}{\mathrm{div}}

\newcommand{\adh}{{\mathrm{adh}}}

\newcommand{\zone}{{01}}
\newcommand{\ztwo}{{02}}
\newcommand{\zk}{{0k}}

\newcommand{\AH}{{A_\mathrm{H}}}
\newcommand{\Areal}{{A_\mathrm{real}}}

\newcommand{\Tn}{T_\mathrm{n}}
\newcommand{\tn}{t_\mathrm{n}}

\newcommand{\bTnk}{{\boldsymbol{T}_{\mathrm{n},k}}}

\newcommand{\btnk}{{\boldsymbol{t}_{\mathrm{n},k}}}

\newcommand{\btt}{\boldsymbol{t}_\mathrm{t}}
\newcommand{\bttk}{{\boldsymbol{t}_{\mathrm{t},k}}}

\newcommand{\ttsmall}{t_\mathrm{t}}
\newcommand{\Ttmax}{T_\mathrm{slide}}
\newcommand{\ttmax}{t_\mathrm{slide}}
\newcommand{\Tnmax}{T_\mathrm{max}}
\newcommand{\gequ}{{g_\mathrm{eq}}}
\newcommand{\gn}{{g_\mathrm{n}}}
\newcommand{\bgn}{{\boldsymbol{g}_\mathrm{n}}}

\newcommand{\gt}{{g_\mathrm{t}}}
\newcommand{\bgt}{{\boldsymbol{g}_\mathrm{t}}}

\newcommand{\gmax}{{g_\mathrm{max}}}

\newcommand{\gcut}{{g_\mathrm{cut}}}
\newcommand{\scut}{{s_\mathrm{cut}}}

\newcommand{\Jcl}{J_{\mathrm{c}\ell}}

\newcommand{\np}{\boldsymbol{n}_\mathrm{p}}

\newcommand{\xk}{{\boldsymbol{x}_k}}

\newcommand{\xlp}{{\boldsymbol{x}_\ell^\mathrm{p}}}

\newcommand {\mra}{\mathrm{a}}
\newcommand {\mrb}{\mathrm{b}}
\newcommand {\mrc}{\mathrm{c}}

\newcommand {\mre}{\mathrm{e}}

\newcommand {\mrm}{\mathrm{m}}
\newcommand {\mrn}{\mathrm{n}}

\newcommand {\mrs}{\mathrm{s}}
\newcommand {\mrt}{\mathrm{t}}

\newcommand {\mrA}{\mathrm{A}}

\newcommand {\mrD}{\mathrm{D}}
\newcommand {\mrE}{\mathrm{E}}

\newcommand {\mrI}{\mathrm{I}}
\newcommand {\mrJ}{\mathrm{J}}

\newcommand {\mrM}{\mathrm{M}}
\newcommand {\mrN}{\mathrm{N}}

\newcommand {\mrP}{\mathrm{P}}

\newcommand {\bzero}{\boldsymbol{0}}

\newcommand {\bff}{\boldsymbol{f}}

\newcommand {\bn}{\boldsymbol{n}}

\newcommand {\bt}{\boldsymbol{t}}
\newcommand {\bu}{\boldsymbol{u}}

\newcommand {\bx}{\boldsymbol{x}}

\newcommand {\bX}{\boldsymbol{X}}

\newcommand {\bsig}{\mbox{\boldmath$\sigma$}}

\newcommand {\sB}{\mathcal{B}}

\newcommand{\AM}{EA}
\newcommand{\AMe}{{\mrE\mrA}}
\newcommand{\CA}{DI}
\newcommand{\CAe}{{\mrD\mrI}}

\newcommand{\Ft}{F_\mrt}
\newcommand{\Fn}{F_\mrn}
\newcommand{\Fzero}{F_0}

\crefname{appendix}{App.}{Apps.}
\Crefname{appendix}{App.}{Apps.}
\Crefname{equation}{Eq.}{Eqs.}
\Crefname{figure}{Fig.}{Figs.}
\crefname{section}{Sect.}{Sects.}
\Crefname{section}{Sect.}{Sects.}
\crefname{table}{Tab.}{Tabs.}
\Crefname{table}{Tab.}{Tabs.}

\hypersetup{colorlinks=true, breaklinks=true, linkcolor=blue, menucolor=blue, citecolor=blue, urlcolor=blue}

\begin{document}


\begin{center}
	\Large{\bf{Continuum contact models for coupled adhesion and friction}}
	\\[1.5ex]
	\large{Janine C. Mergel\footnote{Email:
	mergel@aices.rwth-aachen.de}$^\mra$, Riad Sahli$^\mrb$,
	Julien Scheibert$^\mrb$,
	Roger A. Sauer\footnote{Corresponding author, email:
	sauer@aices.rwth-aachen.de}$^{,\mra}$}
\end{center}

\begin{center}
	\small{$^\mra$\textit{Graduate School AICES, RWTH Aachen University,
		Templergraben 55, 52056 Aachen, Germany}} \\
	\small{$^\mrb$\textit{Laboratoire de Tribologie et Dynamique des
		Syst{\`e}mes LTDS UMR5513, Ecole Centrale de Lyon, ENISE, ENTPE, CNRS,
		F-69134, Ecully, France}} \\
	\vspace{3mm}
	Published in \href{http://dx.doi.org/10.1080/00218464.2018.1479258}
	{The Journal of Adhesion 95(12):1101--1133, 2019} \\
	Submitted on 21 February, 2018; accepted on 17 May, 2018
\end{center}

\vspace{-4mm}
\rule{\linewidth}{.15mm}
{\bf Abstract:} We develop two new continuum contact models for coupled
adhesion and friction, and discuss them in the context of existing models
proposed in the literature. Our new models are able to describe sliding
friction even under tensile normal forces, which seems reasonable for certain
adhesion mechanisms. In contrast, existing continuum models for combined
adhesion and friction typically include sliding friction only if local contact
stresses are compressive. Although such models work well for structures with
sufficiently strong local compression, they fail to capture sliding friction
for soft and compliant systems (like adhesive pads), for which the resistance
to bending is low. This can be overcome with our new models. For further
motivation, we additionally present experimental results for the onset of
sliding of a smooth glass plate on a smooth elastomer cap under low normal
loads. As shown, the findings from these experiments agree well with the
results from our models. In this paper we focus on the motivation and
derivation of our continuum contact models, and provide a corresponding
literature survey. Their implementation in a nonlinear finite element framework
as well as the algorithmic treatment of adhesion and friction will be discussed
in future work.


\vspace{-4mm}
\rule{\linewidth}{.15mm}
\vspace{-8mm}



%



\section{Introduction} \label{s:intro}

This work is concerned with general contact models for adhesion mechanisms that
are characterized by coupled adhesion and friction. Such mechanisms appear in
both natural and technical applications; they also include the adhesive
microstructures of many insects and lizards, which may resemble either soft
adhesive pads (e.g.~for flies and beetles) or rather stiff fibrils (e.g.~for
the gecko). In particular the gecko mechanism has several other remarkable
properties like fast detachment, self-cleaning, and non-clumping
\cite{autumn06c}. To replicate the characteristics of bio-adhesives, various
synthetic materials were developed \cite{geim03, gorb07, parness09,
bartlett12}. Like their natural counterparts, bio-inspired adhesives are often
reusable and removable without leaving any residue. This makes them relevant
for different medical applications; other possible applications include
climbing robots, micro-electro-mechanical systems, and pick-and-place methods.

Initiated by the study of Autumn et al.~\cite{autumn02}, it is generally
believed that bio-adhesion is mainly caused by van der Waals forces.
Nevertheless, also other effects may have an influence, such as
capillary~\cite{huber05b} or electrostatic forces~\cite{izadi14}. Within this
paper, we mainly focus on dry adhesion due to van der Waals interactions. As
experiments reveal, the function of bio- and bio-inspired adhesives further
depends on friction (or ``shear'') forces \cite{autumn06, zhao08}. Thus, the
applications discussed so far do not only involve dry adhesion but also dry
friction. Combined, adhesion and friction play an important role in various
other systems; these include small frictional hairs at the legs of insects
\cite{labonte14}, the peeling of adhesive tapes \cite{amouroux01}, or elastomer
spheres sliding on glass plates \cite{savkoor77, waters10, cohen11, sahli18}.



Regarding bio-adhesive systems in general, experiments show that friction
forces can occur even if the normal force is negative, i.e., tensile
\cite{autumn06, drechsler06, zhao08}. A similar observation can be made in
experiments with chemically modified graphite surfaces \cite{deng12}. Note that
these findings usually refer to the \emph{total} force in the system, while
locally the contact stresses may vary between tension and compression within
the same macroscopic contact area. This was demonstrated by Eason et
al.~\cite{eason15}, who used a tactile sensor to investigate the distribution
of the normal stress underneath an attached gecko toe.

If an adhesive system has a large stiffness (or low compliance, respectively),
its contact surface tends to be characterized by regions of tensile and regions
of compressive contact forces. Then, the fraction of compressive and tensile
areas is large compared to a more compliant system, which may justify the
application of continuum models that capture friction only at the locally
compressed parts of the contact area. However, for compliant, tape-like
structures or soft adhesive pads (e.g.~of insects, see \cite{drechsler06}), the
assumption of a high stiffness is not realistic. Existing continuum models for
combined adhesion and friction include sliding friction only if the local
contact stress is positive, i.e., compressive. Although they work well for
systems with sufficiently strong (local) compression, they fail to capture
sliding resistance for the cases described above.

We thus propose two general contact models for the analysis of physical
problems that are characterized by coupled adhesion and friction. The first
contact model is based on a constant sliding threshold within the contact area,
which means the sliding shear is independent of the local distance of the
contacting surfaces. This model turns out to be favorable for low normal loads,
when friction is purely adhesion-controlled. The second model can be considered
as a local version of Amontons' law extended to adhesion. Both models are
capable of capturing the frictional behavior for zero or negative (tensile)
contact pressures. This approach is motivated by soft and compliant
bio-adhesive pads, which are able to generate friction forces under tensile
normal loads. We additionally present new experimental results for friction
between a smooth elastomer sphere and a smooth glass plate. As shown, the
findings from these experiments agree well with the results for the proposed
models. Our contact models are formulated in the framework of large
deformations, and apply to arbitrary geometries. For many applications this
requires applying computational methods for analysis. An implementation of the
proposed models in the framework of nonlinear finite element methods is
addressed separately \cite{mergelPhD, mergel18comput}.

To summarize, this work provides the following new aspects:
\vspace*{-2ex}
\begin{itemize}\setlength{\itemsep}{0pt}
\item A survey of existing models for dry adhesion and friction, which were
proposed and applied in different fields of research;
\item Experimental results for the onset of sliding of a glass plate on an
elastomer sphere in the presence of low normal loads;
\item Two new continuum models for coupled adhesion and friction of both
biologic and bio-inspired adhesive systems; and
\item A comparison of these models with the experiments.
\end{itemize}



The remainder of this paper is structured as follows. In \cref{s:adhfrict} we
discuss various mechanisms that may contribute to dry adhesion and friction,
and review existing literature. We then present new experimental results in
\cref{s:exp}. In \cref{s:model} we propose two models for adhesion and
friction, and discuss their general behavior. \Cref{s:example} then provides a
comparison of these models with the experiments, for which we consider the
finite element formulation of \cite{mergelPhD, mergel18comput}. \Cref{s:concl}
finally concludes this paper.


\section{Origin and modeling of adhesive friction} \label{s:adhfrict}

This section gives an overview of relevant experimental findings, theoretical
models, and computational approaches that have been proposed in the past. In
the following we will also refer to the terms \emph{sticking} for static
friction, and \emph{sliding} for kinetic (or dynamic) friction, respectively.


\subsection{Adhesive and non-adhesive friction} \label{s:originfrict}

Like adhesion, dry friction may stem from (combinations of) different
mechanisms at smaller length scales. This depends on the involved materials,
surface properties, and the considered length scale. Mechanisms that contribute
to friction are classified in \cite{nosonovsky07} as follows: 1)~adhesion;
2)~deformation of asperities; 3)~plastic deformation at the interface;
4)~fracture; 5)~interlocking; and 6)~wear or contamination particles between
the surfaces. Another possible mechanism is viscous dissipation in the
material, e.g.~for rubber or various tissues. Although sliding friction may be
caused by a mixture of these mechanisms, in engineering applications it is
often described by the well-known formula
\begin{equation}
	\Ft = \mu \, \Fn, \qquad \Fn > 0, \label{e:Coulomb:force}
\end{equation}
where $\mu$ is the coefficient of sliding friction, which relates the normal
and tangential forces, $\Fn$ and $\Ft$, to each other. \Cref{e:Coulomb:force}
results from the following three laws, which were named after Guillaume
Amontons and Charles de Coulomb:
\begin{enumerate}\setlength{\itemsep}{0pt}
	\item[1)] The friction force is proportional to the applied normal load.
		\quad (Amontons' $1\mathrm{st}$ law)\footnote{Note that
		in engineering mechanics, this is often referred to ``Coulomb's law''
		or ``Coulomb friction'' instead.}
	\item[2)] The friction force is independent of the apparent contact area.
		\quad (Amontons' $2\mathrm{nd}$ law)
	\item[3)] The friction force is independent of the sliding
		velocity.\footnote{According to Nosonovsky and Bhushan
		\cite{nosonovsky07}, Coulomb stated this relation for intermediate
		velocities.} \quad (Coulomb's law)
\end{enumerate}
We will refer to relation~\labelcref{e:Coulomb:force} also as
\emph{Coulomb-Amontons friction}. The three laws stated above can be
interpreted by introducing both an apparent (nominal) and a real (true) contact
area,~$\Areal$, and assuming that compressive contact occurs only between small
microasperities. Such an approach goes back to \cite{derjaguin34, bowden39,
archard57}, and explains the apparent contradiction in several experiments
showing the friction force to be proportional to the contact area
\cite{schallamach52, enachescu99}. In many applications, \cref{e:Coulomb:force}
describes frictional sliding of the system appropriately. Nevertheless, for
considerably smooth surfaces or soft materials (such as polymers),
\cref{e:Coulomb:force} may not be valid anymore~\cite{persson08}. This was also
shown for smooth elastomer contacts e.g.~in~\cite{cohen11, sahli18}.

In a study of adhesion and friction of crystal layers,
Derjaguin~\cite{derjaguin34} used statistical analysis to derive a
generalization of Amontons' first law,
\begin{equation}
	\Ft = \mu \, (\Fzero + \Fn) = \mu \, \Areal \, (p_0 + p_\mrn).
	\label{e:extAmontons:orig}
\end{equation}
Here, $p_\mrn := \Fn / \Areal$ is an averaged normal pressure, and $p_0 :=
\Fzero / \Areal$ is an adhesion parameter. After introducing a
material-specific, critical shear stress~$\tau_0 := \mu \, p_0$,
\cref{e:extAmontons:orig} becomes
\begin{equation}
	\Ft = \tau_0 \, \Areal + \mu \, \Fn = \tau \, \Areal, \qquad
	\tau := \tau_0 + \mu \, p_\mrn. \label{e:extAmontons}
\end{equation}
This will be referred to as \emph{extended Amontons' law}. According to Ruths
et al.~\cite{ruths05}, \cref{e:extAmontons:orig} describes the friction force
between dry surfaces that slide smoothly over each other in the presence of
adhesion. For low normal loads, the first term in \cref{e:extAmontons}
predominates, which is called \emph{adhesion-controlled} friction. For large
normal loads, the second term (equivalent to \cref{e:Coulomb:force})
predominates, which is denoted \emph{pressure-controlled} friction.\footnote{In
\cite{ruths05} the expression ``load-controlled'' is used.} The influence of
both terms strongly depends on the specific application, and in particular on
the smoothness, compliance, and adhesion of the contacting surfaces. This is
demonstrated by the experimental results~\cite{homola90} shown in
\cref{f:homola}. For two molecularly smooth (and initially undamaged) mica
cylinders sliding over each other, the friction force is directly proportional
to the contact area; $\Ft = \tau_0 \, \Areal$. This means that with increasing
normal load, the friction force follows the triangles in the figure. At a
sufficiently high normal load, damage occurs and forms wear particles at the
interface. Afterwards, the normal load is decreased again. From that moment on,
the friction force is proportional to the normal load according to
\cref{e:Coulomb:force}, and not to the contact area anymore; see the filled
circles in \cref{f:homola}. As a comparison of \cref{f:homola:1} and
\cref{f:homola:2} reveals, it depends on the specific parameters whether either
adhesion- or pressure-controlled friction provide larger friction forces. The
transition between both is further discussed in~\cite{berman98, gao04b,
jagota11}.
\begin{figure}[h]
	\centering
	\subfigure[Uncoated surfaces.]{
		\begin{minipage}{0.46\textwidth}
		\vspace*{14ex}
		Removed for license reasons from arXiv preprint.
		See Fig.~4(a) in Ref.~\cite{homola90}.
		\vspace*{16ex}
		\end{minipage}
		\label{f:homola:1}
	}\hspace*{0.03\textwidth}
	\subfigure[Surfaces coated with calcium stearate.]{
		\begin{minipage}{0.46\textwidth}
		\vspace*{14ex}
		Removed for license reasons from arXiv preprint.
		See Fig.~4(b) in Ref.~\cite{homola90}.
		\vspace*{16ex}
		\end{minipage}
		\label{f:homola:2}
	}
	\caption{Friction force and contact area measured for two mica cylinders
		sliding over each other in dry air. Dashed arrows indicate the
		transition between adhesion- and pressure-dependent friction.}
	\label{f:homola}
\end{figure}

\Cref{e:extAmontons} has been considered in various research fields, sometimes
with a different terminology. It had already been discussed in a similar form
before Derjaguin~\cite{derjaguin34}, also by Coulomb himself~\cite{gao04b}.
What he proposed was a fracture criterion for granular and brittle materials,
known as the Mohr- or Navier-Coulomb criterion, respectively \cite{deblasio11,
popova15}. Besides, Tabor~\cite{tabor81} used \cref{e:extAmontons} to derive a
pressure-dependent friction model with an effective coefficient of sliding
friction, $\hat{\mu} := \Ft / \Fn = \tau_0/p_\mrn + \mu$. Extended Amontons'
law was further discussed in~\cite{briscoe78, thornton91, yoshizawa93}. It also
appeared in the context of microtribology \cite{briscoe79, berman98b, ruths05},
molecular dynamics simulations \cite{gao04b, sivebaek08, mo09}, and adhesive
friction of bio- and bio-inspired systems \cite{zeng09, hill11}.


\subsection{Existing modeling approaches}\label{s:adhfrict:others}

In the following, we give an overview of the most relevant models for dry
adhesion and friction, with a special focus on continuum methods that are
suitable for computational formulations. For a discussion of friction and
tribology from a physical point of view, we refer to
monographs~\cite{persson00, bhushan13}. In addition, different aspects of
nanotribology are addressed in book~\cite{bhushan05}. A recent review on
modeling and simulation in tribology is given by Vakis et al.~\cite{vakis18}.
For an overview of computational models for adhesion, see also
\cite{sauer16jadh}.


In order to model adhesion (and repulsion) due to van der Waals interactions,
various continuum models have been developed based on the Lennard-Jones (LJ)
potential
\begin{equation}
	\phi(r) = \varepsilon \, {\Big(\frac{r_0}{r}\Big)}^{12}
	- 2\,\varepsilon \, {\Big(\frac{r_0}{r}\Big)}^6. \label{e:LJ}
\end{equation}
Here, $\phi$ is the potential between two molecules separated by the
distance~$r$; it depends on the parameters~$\varepsilon$ and $r_0$, which
denote the depth of the potential well and the molecular equilibrium distance,
respectively. In order to increase the efficiency of computational models based
on \cref{e:LJ}, the interactions between two (in general arbitrarily shaped)
bodies are often approximated by effective volumetric forces \big(with unit
$\mrN/\mrm^3$\big) or effective surface tractions \big(with unit
$\mrN/\mrm^2$\big). Such an approach is pursued in various computational
formulations \cite{sauer07ijnme, du07, sauer09cmame, zhang11, fan15a,
jayadeep16}.

In general, models based on the Lennard-Jones potential do not contain any
contribution due to friction. For this reason, Deng et al.~\cite{deng12}
additionally consider a constant tangential contact stress during sliding
wherever the normal contact stress is compressive. This corresponds to the
first term in \cref{e:extAmontons}, and thus shows similarities to one of our
proposed models (\cref{s:model:CA}). Apart from that, Jiang and
Park~\cite{jiang15} propose to extend the original Lennard-Jones potential by
an additional ``friction potential'' in order to describe the friction
properties of layered materials such as graphene. However, this approach is
restricted to static friction.


Van der Waals forces can also be modeled by fitting traction-separation laws
within a cohesive zone model; see e.g.~\cite{yao06a, peng10, sauer13cmbbe,
sauer14finel}. Although these papers use models that do not include tangential
sliding, there also exist numerous cohesive zone models that additionally
include sliding friction. In early works \cite{tvergaard90, lissenden95},
sliding friction according to Coulomb and Amontons was modeled subsequently
after full tangential debonding. More recent cohesive zone models account for
sliding friction also during partial separation \cite{chaboche97b, raous02,
delpiero10, snozzi13}. Besides, there exist several cohesive zone models
including friction, which are based on representative interface or volume
elements \cite{alfano06, sacco10, sacco12, guiamatsia14}.

In all of the cohesion models discussed so far, the damage caused by debonding
is considered to be irreversible. This, however, is not realistic for van der
Waals interactions. Schryve~\cite{schryvePhD} and Cocou et al.~\cite{cocou10}
thus modified model~\cite{raous99} to capture partial recovering (``healing'')
of the adhesive strength when the gap between the interfaces is closed again.
Schryve applied this approach to investigate adhesion and friction between
elastomer and glass.


Macroscopic adhesion and friction can also be modeled by statistical
homogenization of the microscopic contact behavior \cite{wriggers06}. Tworzydlo
et al.~\cite{tworzydlo98}, for instance, combined finite element simulations of
small surface asperities with statistical analysis, mainly focusing on metallic
surfaces. In that study adhesion was modeled with an integrated Lennard-Jones
potential. Besides, there exists a recursive multiscale
approach~\cite{wriggers09} to investigate contact of rubber and rough, rigid
surfaces (like tire-road contact). This model captures van der Waals
interactions by means of a fitted cohesive zone model, and restores adhesion
fully when contact is re-established.


For completeness we also outline the most relevant models for non-adhesive
friction. In a general continuum framework, sliding
law~\labelcref{e:Coulomb:force} is usually stated in terms of the normal and
tangential contact tractions, $\tn$ and $\btt$,
\begin{equation}
	\|\btt\| = \mu \, \tn, \qquad \tn > 0. \label{e:Coulomb:tract}
\end{equation}
In many engineering applications it is sufficient to consider the coefficient
of sliding friction, $\mu$,~as a constant parameter for the material pair at
the interface; see e.g.~\cite{ciavarella98, tromborg11, amundsen15,
yastrebov16}. Depending on the application, however, $\mu$ may be affected by
the sliding velocity \cite{grosch63, wubavouzet10}, contact pressure
\cite{scheibert08}, temperature \cite{grosch63}, or microscopic time scales
\cite{tromborg14}. For an overview of approaches suitable for different kinds
of materials see~\cite{wriggers06}.

Note that both the friction and the cohesive zone models discussed above
include sliding friction only under local compression. This means that after
tangential debonding, frictionless sliding occurs if the normal contact
stresses are tensile.


Next, we outline the most relevant small-deformation models for adhesion and
friction. Regarding the frictionless adhesion of spheres, these include the
well-known Johnson-Kendall-Roberts (JKR) \cite{johnson71} and
Derjaguin-Muller-Toporov (DMT) \cite{derjaguin75} models, which apply to
compliant and stiff materials, respectively. Both models were later combined
in~\cite{maugis92}. Although these models are restricted to small deformations
and special contact geometries, they have been successfully applied to fit
experimental data. In the experimental results shown in \cref{f:homola}, for
instance, the contact area agrees well with that predicted by the JKR theory.
For adhesion-controlled friction, it is then possible to determine the
frictional shear stress~$\tau_0$ (a material-pair parameter) from sphere-plane
contact as follows: 1)~Predict the real contact area, $\Areal$, with the JKR
theory by inserting Young's modulus, the sphere radius, the applied normal
load, and the interfacial adhesion energy; 2)~measure the friction force,
$\Ft$; and 3)~combine both values using $\Ft = \tau_0 \, \Areal$; see also
\cref{s:exp}. In Savkoor and Briggs~\cite{savkoor77}, the JKR theory was
extended further to incorporate the influence of tangential shearing. According
to that model, for a fixed normal force, the contact area decreases with
increasing tangential force. This agrees with the experimental findings
presented therein. Related studies and important extensions
include~\cite{johnson96, johnson97, waters10}.

As mentioned in \cref{s:originfrict}, friction may be explained by small
asperities sliding over each other. Asperity-based friction models
\cite{braun08, thogersen14} generally differ in the surface shape and the
number of involved length scales; examples include periodic and wavy substrate,
fractal surfaces, or stochastic profiles. Note that such approaches are also
used in many of the papers outlined in \cref{s:originfrict}, in particular to
verify a linear relation between the real contact area and the normal force
(see also \cite{majumdar91, persson01, barber13}). Asperity-based models are
not only used to investigate friction, but also adhesion. Multiasperity contact
was combined both with the JKR \cite{fuller75} and DMT theory \cite{chang88,
maugis96}. Further extensions include~\cite{greenwood81, rabinovich00}. Like
asperity-based friction models, models for rough surface adhesion can be based
e.g.~on a single layer of wavy asperities or on single- and multiscale fractal
surfaces, respectively \cite{pastewka14, mueser17}. For an overview see
also~\cite{persson05}.

Finally, there exists the analytical tape model~\cite{kendall75} by Kendall,
which was also extended numerous times to investigate directional peeling of
thin strips or tapes. Some of these models also incorporate sliding friction.
As experiments show, approaches of this type are suitable to accurately capture
the peeling force of structures with small bending stiffness such as elastomer
films \cite{collino14}, adhesive tape \cite{dalbe16}, and adhesive beetle pads
\cite{labonte16}. These models are thus often applied to discuss the properties
of bio-adhesive systems. For adhesive fibrils with finite bending stiffness,
however, such models may underestimate the actual peel-off force
considerably~\cite{sauer11ja, mergel14jadh}.


\section{Experimental results for sliding of a smooth elastomer sphere}
\label{s:exp}

\begin{figure}[h]
	\unitlength\textwidth
	\begin{picture}(1,0.35)
		\put(0,0){\includegraphics[width=0.65\textwidth]{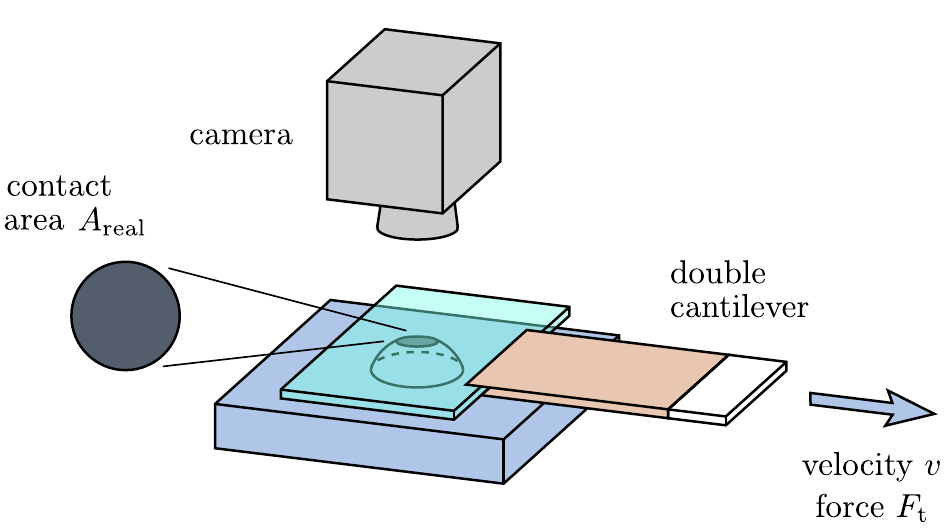}}
		\put(0.50,0.24){\small \begin{tabular}{r@{\quad}c@{\ \ }l}
			sphere: & $\bullet$ & material: PDMS Sylgard 184 \\[0.3ex]
			& $\bullet$ & Young's modulus: $E = 1.6 \pm 0.1\,\mrM\mrP\mra$
				\\[0.3ex]
			& $\bullet$ & Poisson's ratio: $0.5$ \\[0.3ex]
			& $\bullet$ & radius of curvature: $9.42\,\mrm\mrm$ \\[0.3ex]
			& $\bullet$ & diameter of cap: $12\,\mrm\mrm$ \\[0.3ex]
			& $\bullet$ & work of adhesion (PDMS / glass): \\[0.3ex]
			&& $\Delta\gamma = 27 \pm 1\,\mrm\mrJ/\mrm^2$
		\end{tabular}}
	\end{picture}
	\caption{Monitoring of the incipient sliding of a glass plate on an
		elastomer sphere: Sketch of the experimental setup.}
	\label{f:exp:setup}
\end{figure}

We now present new experimental results for the onset of sliding of a smooth
glass plate on a smooth polydimethylsiloxane (PDMS) sphere under low normal
loads. Our experimental setup, inspired by Prevost et al.~\cite{prevost13} and
fully described in Sahli et al.~\cite{sahli18}, is illustrated in
\cref{f:exp:setup}. As shown, the glass plate is attached to a double
cantilever, which can deflect in the vertical direction (vertical stiffness
$410 \pm 20\,\mrN/\mrm$). The other end of the cantilever is mounted on a
vertical translation stage (not shown in the figure), which allows the glass
plate to be brought into and out of contact with the elastomer sphere. First
contact is found by very slowly moving the double cantilever down and stopping
as soon as the sphere snaps into adhesive contact with the glass. Going further
down allows to obtain interfaces with increasing initial contact areas,
$\Areal$. By analyzing the reflection images of the contact interface (see the
method described in \cite{sahli18}), it is possible to monitor the evolution
of~$\Areal$ over time. The vertical displacement is then kept constant during
the experiments.

Interestingly, once first contact has been formed, it is possible to move the
double cantilever back up some distance without losing contact. This allows us
to obtain interfaces in a global tensile state. Although our setup does not
allow for measuring the resulting normal load directly, it is possible to
estimate it from the initial contact area using the JKR
theory~\cite{johnson71}. This had been confirmed in an initial calibration on a
dedicated device considering the same materials and contact geometry: Fitting
the data by means of the JKR formula provided the values for Young's modulus of
our PDMS as well as the adhesion energy of our glass/PDMS interface (see
\cref{f:exp:setup}).
\begin{figure}[h]
	\centering
	\subfigure[Normal load vs.~measured initial contact area.]{
		\includegraphics[width=0.47\textwidth]
			{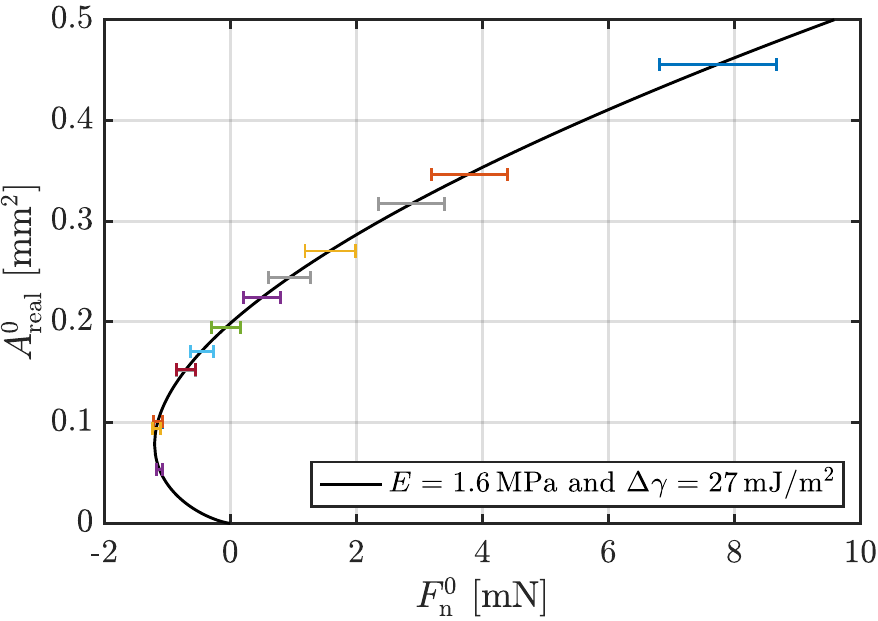} \label{f:exp:FnJKR:Ac}
	}\hspace*{1ex}
	\subfigure[Normal load vs.~contact radius $r_\mathrm{real}^0
		= \sqrt{A_\mathrm{real}^0/\pi}$.]{
		\includegraphics[width=0.47\textwidth]
			{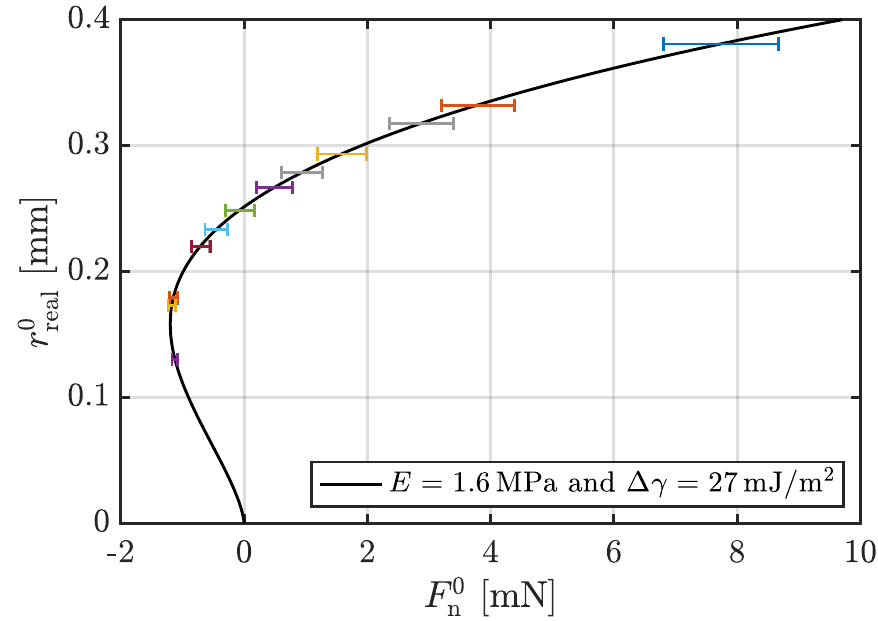} \label{f:exp:FnJKR:rc}
	}
	\caption{Initial contact of a glass plate on an elastomer sphere: Initial
		normal load~$F_\mrn^0$ estimated from the measured initial contact area
		using the JKR theory and the material parameters in \cref{f:exp:setup};
		each colored bar corresponds to a different experiment.}
	\label{f:exp:FnJKR}
\end{figure}
\Cref{f:exp:FnJKR:Ac} shows the range of the initial normal forces~$F_\mrn^0$
in dependence of the initial contact areas~$A_\mathrm{real}^0$ measured in our
experiments. \Cref{f:exp:FnJKR:rc} shows an alternative representation based on
the corresponding contact radii. The horizontal bars in \cref{f:exp:FnJKR}
indicate the possible range of values caused by uncertainties in the material
parameters~$E$ and~$\Delta \gamma$ (\cref{f:exp:setup}). As these results
imply, for some of the experiments the initial normal load is either close to
zero or tensile.

\begin{figure}[h]
	\unitlength\textwidth
	\centering
	\begin{picture}(0.995,0.155)
		\put(0,0){\includegraphics[width=0.125\textwidth]{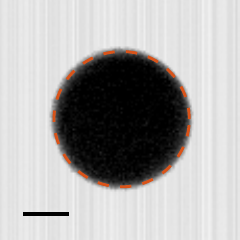}}
		\put(0.145,0){\includegraphics[width=0.125\textwidth]{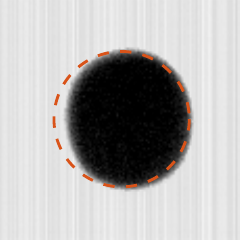}}
		\put(0.29,0){\includegraphics[width=0.125\textwidth]{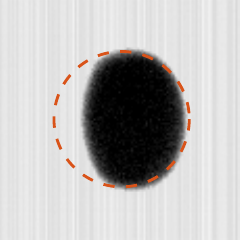}}
		\put(0.435,0){\includegraphics[width=0.125\textwidth]{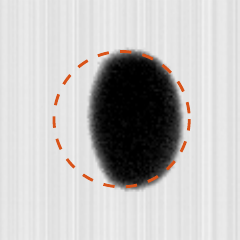}}
		\put(0.58,0){\includegraphics[width=0.125\textwidth]{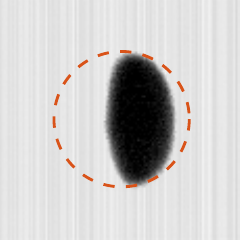}}
		\put(0.725,0){\includegraphics[width=0.125\textwidth]{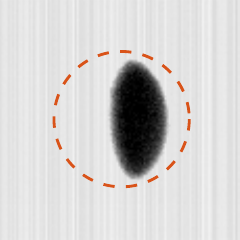}}
		\put(0.87,0){\includegraphics[width=0.125\textwidth]{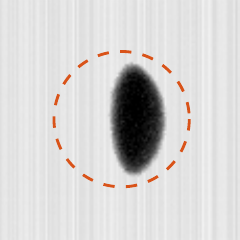}}
		\put(0.035,0.14){\small $0.00\,\mrs$}
		\put(0.180,0.14){\small $0.45\,\mrs$}
		\put(0.325,0.14){\small $0.62\,\mrs$}
		\put(0.470,0.14){\small $0.73\,\mrs$}
		\put(0.615,0.14){\small $0.88\,\mrs$}
		\put(0.760,0.14){\small $1.25\,\mrs$}
		\put(0.905,0.14){\small $1.75\,\mrs$}
	\end{picture}
	\caption{Incipient sliding of a glass plate on an elastomer sphere:
		Snapshots of the contact area for an initial area of $0.27 \,
		\mrm\mrm^2$, which is the upper yellow curve in \cref{f:exp:AcTime};
		the sphere is fixed with respect to the frame of the camera, while the
		glass plate moves to the right; the driving velocity is
		$v = 0.1\,\mrm\mrm/\mrs$; the bar shown in the first figure corresponds
		to a length of $0.25\,\mrm\mrm$.}
	\label{f:exp:snapshots}
\end{figure}
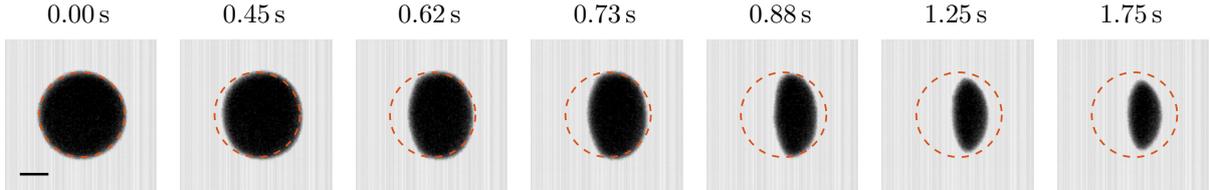

The vertical translation stage at the end of the double cantilever is mounted
on a motorized horizontal translation stage, enabling motion of the glass plate
at a constant velocity $v = 0.1 \,\mrm\mrm/\mrs$. We measure the time evolution
of the tangential load, $\Ft$, by means of a load sensor (resolution
ca.~$1\,\mrm\mrN$) as well as the time evolution of~$\Areal$. A typical
evolution of the contact area is shown in the snapshots of
\cref{f:exp:snapshots}. As one can see, with increasing tangential loading, the
area shrinks due to its left and right edges moving towards each other (with
respect to the frame of the camera and the base of the sphere), while the left
contact edge starts moving first.

\begin{figure}[ht]
	\centering
	\subfigure[Tangential force over time.]{
		\includegraphics[width=0.47\textwidth]{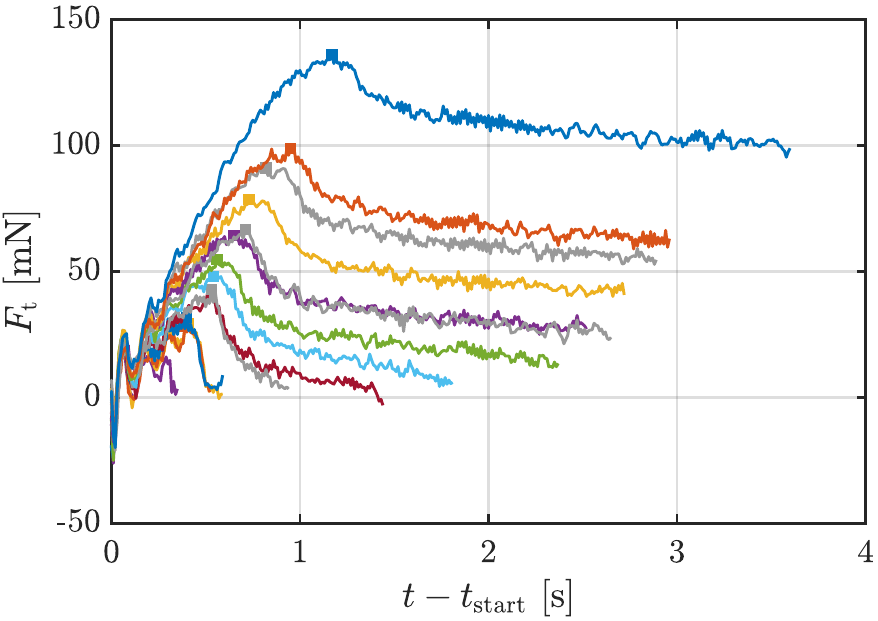}
		\label{f:exp:FtTime}
	}\hspace*{1ex}
	\subfigure[Contact area over time.]{
		\includegraphics[width=0.47\textwidth]{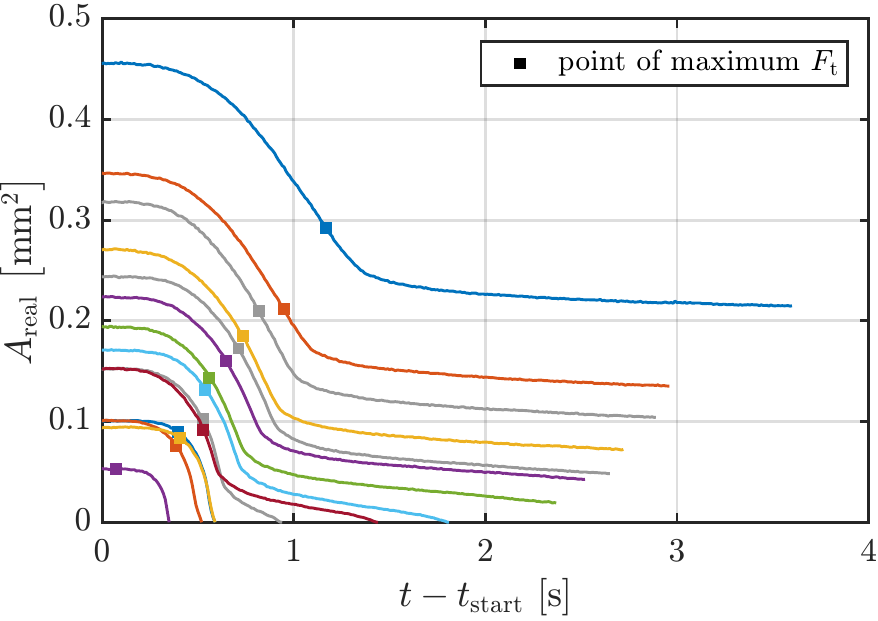}
		\label{f:exp:AcTime}
	}
	\subfigure[Contact area as a function of the tangential force.]{
		\includegraphics[width=0.47\textwidth]{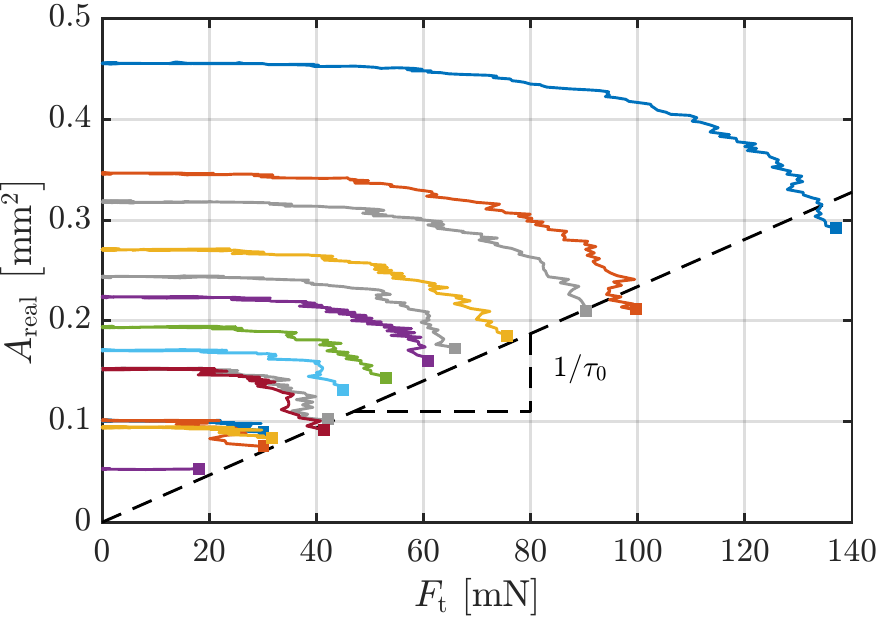}
		\label{f:exp:AcFt}
	}\hspace*{1ex}
	\subfigure[$\Ft\,/\,\Areal$ over time for the largest initial contact.]{
		\includegraphics[width=0.47\textwidth]{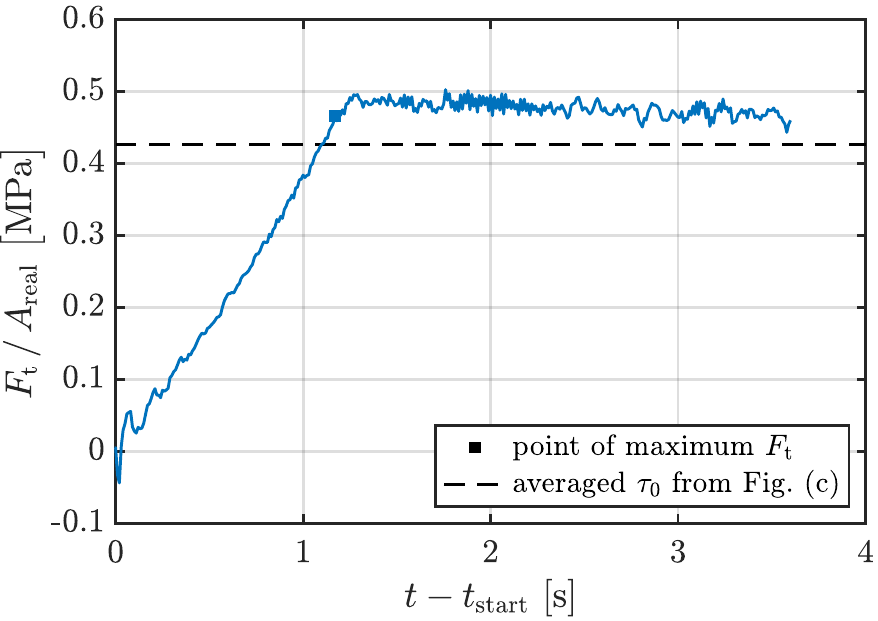}
		\label{f:exp:tauTime}
	}
	\caption{Experimental results for the incipient sliding of a glass plate on
		an elastomer sphere: Friction force~$\Ft$ and contact area $\Areal$
		measured under low normal loads; each color corresponds to a different
		initial contact area (also shown in \cref{f:exp:FnJKR}); the squares
		indicate the states with maximum $\Ft$; $t_\mathrm{start}$ is the
		instant when the motor starts to move at velocity
		$v = 0.1\,\mrm\mrm/\mrs$.}
	\label{f:exp:results}
\end{figure}

\Cref{f:exp:FtTime} shows the tangential force, $\Ft$, as a function of time
for different initial contact areas. Here, the origin of time,
$t_\mathrm{start}$, is taken when the motor starts moving. For each curve, the
typical behavior is the following: $\Ft$ increases, first almost linearly and
then with a weakening slope, reaches its maximum (the static friction peak),
and rapidly drops afterwards before entering a slow decay during macroscopic
sliding. This slow decay, arising from a small residual angle between the glass
plate and the horizontal, was negligible in~\cite{sahli18}. Here, it is
detectable because of the (about 20 times) smaller initial areas and the (about
8 times) stiffer cantilever.

The time evolution of the contact area $\Areal$ is plotted in
\cref{f:exp:AcTime}. As shown also in~\cite{savkoor77, waters10, sahli18},
$\Areal$~decreases as the interface is progressively sheared. The rate of area
decrease (i.e., the slope of the curve) significantly drops when the contact
enters macroscopic sliding. The subsequent slow decrease of the area is the
counterpart of the slow decrease of the friction force. Consistently with
Waters and Guduru~\cite{waters10}, contacts with the smallest initial areas
(below about $0.1\,\mrm\mrm^2$) abruptly vanish upon shearing, without entering
a macroscopic sliding regime. In contrast, the contacts with the highest
initial areas (above about $0.19\,\mrm\mrm^2$) do not vanish during the time
window shown in \cref{f:exp:FtTime,f:exp:AcTime}. For each experiment, this
time window covers the time that is necessary to reach the static friction peak
(squares in \cref{f:exp:FtTime}), and to slide further by a distance that is
equal to half times the width of the contact area at the static friction peak.

\Cref{f:exp:AcFt} shows the dependency between the contact area and the
tangential force in the incipient loading phase, before the friction peak is
reached. As can be seen, at the onset of sliding (see the squares) the friction
force and the contact area are, to a good approximation, proportional to each
other. This is consistent with experimental results on identical smooth
sphere/plane contacts~\cite{sahli18}, albeit for much larger contacts. It is
also consistent with experiments on rough contacts involving soft materials
(see e.g.~\cite{wubavouzet10, degrandicontraires12, yashima15, sahli18}).

\Cref{f:exp:tauTime} exemplarily shows the evolution of the ratio
$\Ft\,/\,\Areal$ over time for the largest initial contact area ($\Areal =
0.46\,\mrm\mrm^2$, upper dark blue curve in
\crefrange{f:exp:FtTime}{f:exp:AcFt}), for which both the small residual
inclination of the glass plate and the resolution of the force sensor have
lowest influence. This ratio corresponds to the average tangential traction
within the contact area. As seen, it first increases almost linearly until the
maximum friction force is reached. From then on, the ratio $\Ft\,/\,\Areal$
remains nearly constant. As these results demonstrate, considering a constant
shear stress during sliding is a reasonable assumption when modeling friction
of soft and smooth surfaces. This corresponds to adhesion-controlled friction,
when $\Ft \approx \tau_0\, \Areal$ according to \cref{e:extAmontons}. For our
experiments, we measure an average value of $\tau_0 = 0.43 \pm
0.01\,\mrM\mrP\mra$, which is also visualized in \cref{f:exp:AcFt}.


\section{Models for adhesive friction}\label{s:model}

Before formulating adhesive friction between two arbitrary objects
mathematically, we must first quantify the separation of their surfaces
appropriately. To this end, we conceptually introduce variables for the
distances in normal and tangential directions following classical contact
formulations \cite{laursen02,wriggers06}. Using a common notation in continuum
mechanics (see e.g.~\cite{bonet97}), we use uppercase letters for variables
defined in the \emph{reference} configuration of a body, and small letters for
variables in the \emph{current} configuration.

For a given point~$\xk$ on the contact surface of one of the two bodies
$\sB_k$, $k=1,2$, we first introduce its closest neighbor (or
\textit{projection point}),~$\xlp$, on the contact surface of the opposing body
$\sB_\ell$, $\ell=2,1$. At~$\xlp$, the outward unit normal vector of the
surface is denoted~$\np$. Once these quantities are determined, we define a
normal gap vector, $\bgn$, as well as a (scalar) signed normal gap,~$\gn$, as
\begin{equation}
	\bgn = \xk - \xlp, \qquad \gn = \bgn \cdot \np. \label{e:gn}
\end{equation}
In addition, we conceptually introduce a vector for the tangential slip,
$\bgt$, which reduces to the scalar~$\gt$ for 2D problems. This quantity
contains the magnitude and direction of the tangential displacement between the
two surfaces. The normal and tangential gaps, $\gn$ and $\bgt$, as well as the
normal vector, $\np$, will be used to derive and illustrate our new contact
models. Their computation is addressed e.g.~in~\cite{sauer13cmame,
sauer15ijnme}, while the algorithmic treatment of combined adhesion and
friction is discussed by Mergel et al.~\cite{mergelPhD, mergel18comput}.


\subsection{Adhesive and repulsive contact}\label{s:model:CGCM}

\begin{figure}[ht]
	\centering
	\includegraphics[width=0.47\textwidth]{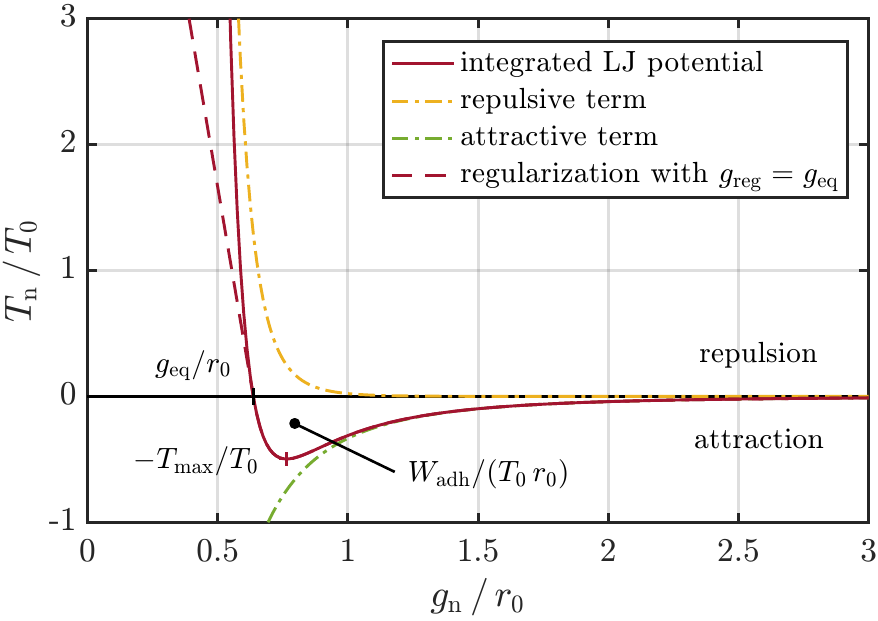}
	\caption{Contact traction in the model of Sauer and
		Wriggers~\cite{sauer09cmame} (``surface force formulation'') for
		frictionless adhesion and repulsion; $T_0 = \AH\,/\,(2\pi r_0^3)$.}
	\label{f:Tn}
\end{figure}

In order to describe general adhesion and repulsion between two bodies under
large deformations, we consider the coarse-grained contact model (CGCM)
\cite{sauerPhD, sauer07ijnme, sauer09cmame}. This model is derived by first
integrating the Lennard-Jones (LJ) potential from \cref{e:LJ} over one of the
two bodies (index~$\ell$), assuming that its surface appears flat in the region
affected by~$\phi$ (typically a few nanometers). This yields a volumetric force
\big(with unit $\mrN/\mrm^3$\big), which acts at each point within the other
body (index~$k$). In a second step, this body force is projected onto the
surface of body~$\sB_k$ to obtain the contact traction \big(with unit
$\mrN/\mrm^2$\big)
\begin{equation}
	\bTnk = \frac{\theta_k}{J_\ell} \, \Tn(\gn) \, \np, \qquad
	\Tn(\gn) = \frac{\AH}{2\pi r_0^3} \left[{ \frac{1}{45}
	{\Big(\frac{r_0}{\gn}\Big)}^9 - \frac{1}{3}
	{\Big(\frac{r_0}{\gn}\Big)}^3}\right]. \label{e:Tn}
\end{equation}
Note that this contact law is stated in the undeformed \emph{reference}
configuration. As can be seen in \cref{f:Tn}, $\Tn$ can be split into a
repulsive (power~9) and an attractive (power~3) term. In \cref{e:Tn}, the
parameter $\AH = 2\pi^2 \, \beta_\zone \, \beta_\ztwo \, \varepsilon \, r_0^6$
is the Hamaker constant \cite{israelachvili11}, which contains the initial
molecular densities~$\beta_\zk$ of the two bodies and the
parameters~$\varepsilon$ and $r_0$ of the Lennard-Jones potential. The scalar
$J_\ell \geq 0$ denotes the volume change of the surrounding material during
deformation. It corresponds to the determinant of the deformation gradient of
the body~$\sB_\ell$ \cite{sauer09cmame}. If this body is rigid, incompressible,
or considerably stiffer than body~$k$, one can assume that $J_\ell$ is either
equal or close to one. The volume change can be related to the local surface
stretch~$\Jcl$ by using $J_\ell = \Jcl \cdot \lambda_\ell$, where
$\lambda_\ell$ is the stretch along the thickness (perpendicular to the
surface). Since the potential~$\phi(r)$ rapidly decays to zero at large
separations, it is reasonable to assume that within its effective range
$\lambda_\ell \approx 1$, see also Sect.~2.3.2 of~\cite{sauer13cmame}.

\cref{e:Tn} contains a scalar, $\theta_k$, that includes the current alignment
of the two interacting surfaces as well as the volume change of body~$\sB_k$.
Like~$J_\ell$, $\theta_k$ requires the computation of the deformation in the
vicinity of surface point $\bx_k$. If both surfaces are parallel, and if the
influence of the surface stretch of body~$\sB_k$ is negligible, one can set
$\theta_k \approx 1$~\cite{sauer09cmame}. In the following, we also assume that
$\theta_k \approx 1$. An alternative approximation (similar to $J_\ell \approx
\Jcl$) is proposed by Mergel in~\cite{mergelPhD}.

Before we proceed with the frictional part, let us define some characteristic
parameters also shown in \cref{f:Tn}:
\begin{enumerate}\setlength{\itemsep}{0pt}
\item The equilibrium distance $\gequ = r_0\,/\,\sqrt[6]{15}$ at which
	$\Tn(\gequ) = 0$ \cite{sauer11ja};
\item The work of adhesion per unit area for full separation, which is obtained
	by integrating $|\Tn(\gn)|$ from $\gn = \gequ$ to $\gn = \infty$
	\cite{sauer11ja},
	\begin{equation}
		W_\adh = \frac{\sqrt[3]{15} \, \AH}{16 \pi r_0^2}; \label{e:Wadh}
	\end{equation}
\item The location~$\gmax$ of the maximum adhesive traction~$\Tnmax$,
	\begin{equation}
		\gmax = \frac{r_0}{\sqrt[6]{5}}, \qquad
		\Tnmax = |\Tn(\gmax)| = \frac{\sqrt{5}\,\AH}{9\pi r_0^3}.
		\label{e:gmaxTnmax}
	\end{equation}
	Since $\Tn(\gmax)$ is negative (i.e., attractive), $\gmax$ is the
	global minimum of $\Tn$.
\end{enumerate}
Furthermore, in a computational implementation, one may increase robustness by
regularizing the normal traction for small normal gaps; see
\cref{a:regularize,f:Tn}. This approach prevents ill-conditioning caused by the
slope of~$\Tn$ approaching minus infinity for decreasing~$\gn$.


\subsection{Frictional contact}\label{s:model:frict}

We now propose two new phenomenological models that combine adhesion and
repulsion with sliding friction. To this end, we assume that the sliding
resistance is equal to the threshold for static friction. This means that the
tangential traction required to initiate the sliding process agrees with the
traction in the final sliding state. This assumption is physically justified by
experimental observations for both biological adhesives and rough elastomers,
indicating that for such systems static friction is comparable to kinetic
friction \cite{autumn00, autumn06, zhao08, gravish10, prevost13}. For the
validity and restrictions of this assumption see also \cref{s:model:validity}.

To shorten the notation, we now omit index~$k$. Let $\btt(\gn,\bgt)$ denote the
tangential traction vector due to frictional sticking or sliding, which is a
force per \emph{current} area. $\btt$~depends on both the normal gap from
\cref{e:gn} and the tangential slip~$\bgt$. We now assume that~$\btt$ satisfies
\begin{equation}
	\|\btt(\gn,\bgt)\|\ \begin{cases}
	\ <\ \ttmax(\gn)	& \text{during\ sticking}, \\
	\ =\ \ttmax(\gn)	& \text{during\ sliding},
	\end{cases} \label{e:ttorig}
\end{equation}
where $\ttmax(\gn) \ge 0$ is a function that defines the sliding threshold.
For~$\ttmax$ we propose two different approaches in the next two subsections.
Note that the classical Coulomb-Amontons law in \cref{e:Coulomb:tract}
corresponds to $\ttmax := \mu \, \tn$ for $\tn > 0$.


\subsubsection{Model~\CA: Distance-independent sliding friction in the contact
area} \label{s:model:CA}

\begin{figure}[h]
	\centering
	\subfigure[Sliding threshold for varying $\gn$.]{
		\includegraphics[width=0.47\textwidth]{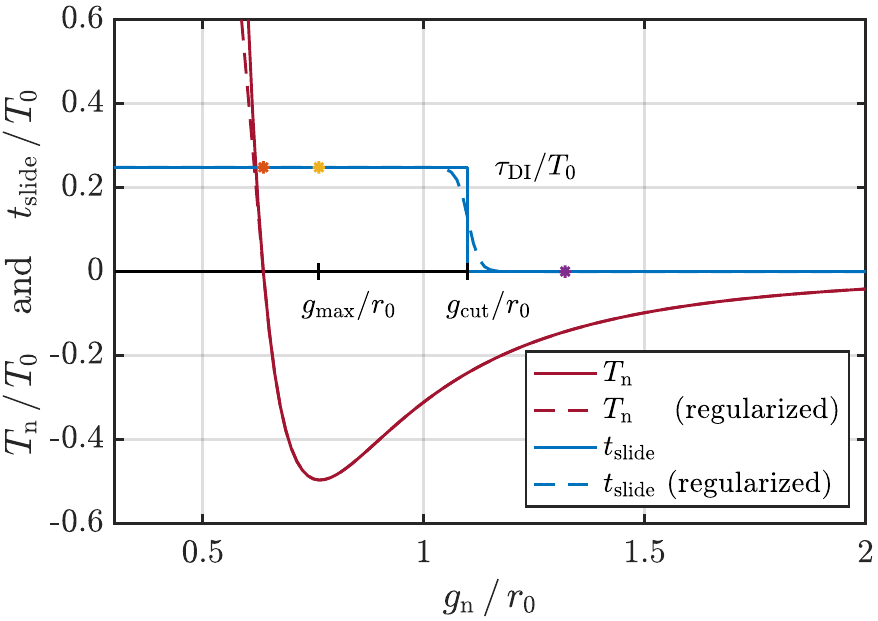}
		\label{f:tslideCA}
	}\hspace*{1ex}
	\subfigure[Friction law in 2D for different but fixed $\gn$.]{
		\includegraphics[width=0.47\textwidth]{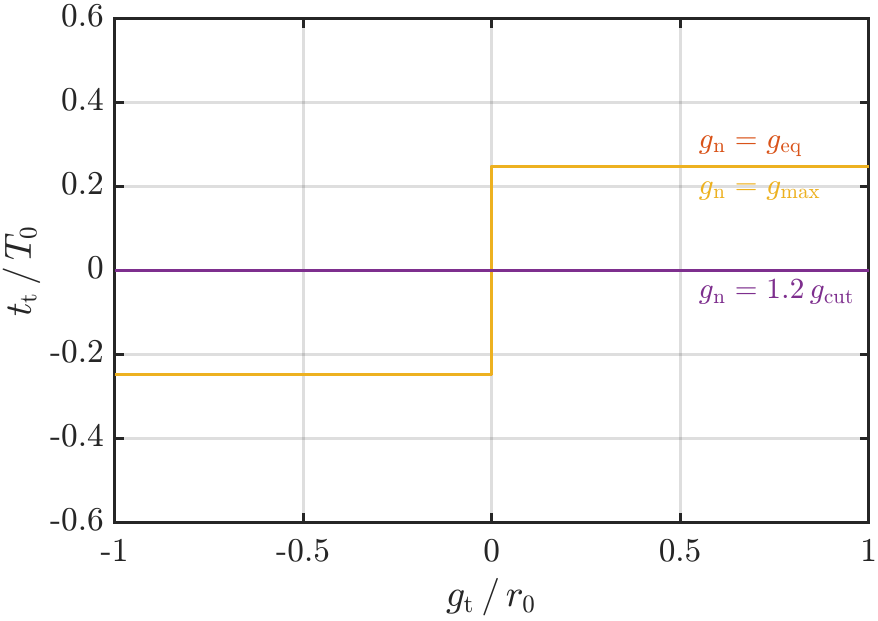}
		\label{f:ttlawCA}
	}
	\caption{Model~\CA: Constant sliding traction within the contact area
		defined by the cutoff distance~$\gcut$; $\tau_\CAe = 0.5 \, \Tnmax$,
		$\gcut = 1.1\,r_0$, and $T_0 = \AH \,/\, (2\pi r_0^3)$; the colored
		asterisks mark the normal gap values $\gequ$~(where $\Tn = 0$, orange),
		$\gmax$~(yellow), and $1.2\,\gcut$ (purple).}
	\label{f:modelCA}
\end{figure}

The first proposed friction law is motivated by the experimental results
discussed in \cref{s:exp}. We simply assume that the sliding threshold is
independent of the distance~$\gn$, i.e., constant within the current contact
area. As a consequence, the resulting friction force is proportional to this
area if the entire body is sliding. Let us first define some cutoff distance,
$\gcut$, up to which the surfaces are sufficiently close to each other in order
to experience friction. After introducing a constant parameter~$\tau_\CAe > 0$
for the frictional shear strength, we define
\begin{equation}
	\ttmax(\gn) = \begin{cases}
		\tau_\CAe,	& \gn \le \gcut, \\
		0,			& \gn > \gcut.
	\end{cases} \label{e:tslideCAunreg}
\end{equation}
Since this function is discontinuous at $\gn = \gcut$, we regularize it with
the logistic function,
\begin{equation}
	\ttmax(\gn) = \frac{\tau_\CAe}{1 + \mre^{\, k_\CAe (\gn - \gcut)}},
	\label{e:tslideCA}
\end{equation}
where $k_\CAe > 0$ (with unit $1/\mrm$) is a sufficiently large parameter. Both
the original and the regularized models are illustrated in \cref{f:tslideCA}.
\Cref{f:ttlawCA} depicts the resulting friction law, i.e., the tangential
traction-separation relation for arbitrary but fixed normal distances.

Note that for the regularized model $\ttmax(\gcut) = \tau_\CAe/2$. As seen in
\cref{f:modelCA}, for model~\CA\ both the value~$\tau_\CAe$ and the cutoff
distance~$\gcut$ can be chosen independently of the normal traction.
Nevertheless, for the sake of comparison with the other model we introduce a
coefficient that relates the constant~$\tau_\CAe$ to the maximum adhesive
traction~$\Tnmax$, defined in \cref{e:gmaxTnmax}, as
\begin{equation}
	\mu_\CAe := \frac{\tau_\CAe}{\Tnmax}. \label{e:muCA}
\end{equation}
The sliding threshold $\tau_\CAe$ is related to the \emph{actual} contact
surface (in the \emph{current} configuration). This is motivated by the
experiments from \cref{s:exp}, for which the force is proportional to the
current size of the contact area. If we defined the parameter~$\tau_\CAe$ in
the \emph{reference} configuration instead, we would miss the change in the
contact area. The differences between those two approaches are discussed
in~\cite{mergelPhD}.

When recapitulating the unregularized version of model~\CA, one may recognize
that \cref{e:tslideCAunreg} has some similarities to one of the earliest
cohesive zone models, the Dugdale model~\cite{dugdale60}. In contrast to that
model, however, we here define the sliding threshold for dynamic friction
(instead of the normal stress during pure debonding). As mentioned also in
\cref{s:adhfrict:others}, a similar approach is used in~\cite{deng12} to model
sliding of graphene sheets. Therein, a constant stress during sliding is
considered for those parts of the surfaces that are in compressive contact.
This would correspond to the unregularized friction
law~\labelcref{e:tslideCAunreg} with $\gcut = \gequ$. The current model is more
general, because it can also be used to describe sliding resistance for tensile
contact.


\subsubsection{Model~\AM: Extended Amontons' law in local
form}\label{s:model:AM}

The second proposed traction-separation law is inspired by extended Amontons'
law~\labelcref{e:extAmontons}. Importantly, that law was originally formulated
in terms of force resultants and average tractions. In our continuum
formulation, however, the normal contact stress~\labelcref{e:Tn} can vary
within the contact area between attraction and repulsion. To include a
tangential resistance against sliding even for zero or negative normal
pressures, we now propose a model that can be regarded as extended Amontons'
law in \textit{local} form.

\begin{figure}[p]
	\centering
	\subfigure[Sliding threshold for varying $\gn$; $\scut = 0$.]{
		\includegraphics[width=0.47\textwidth]{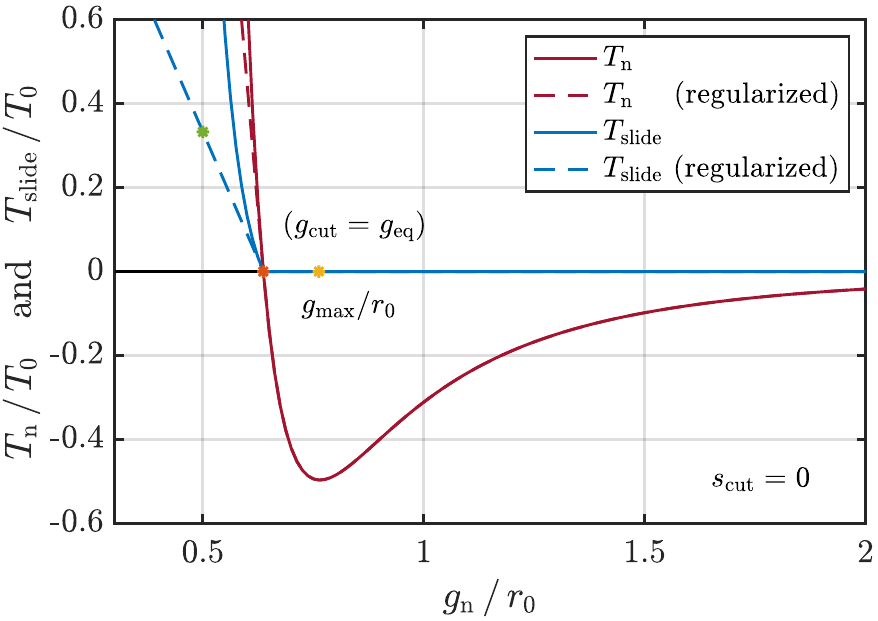}
		\label{f:TslideAM:s0}
	}\hspace*{1ex}
	\subfigure[Friction law for different but fixed $\gn$; $\scut = 0$.]{
		\includegraphics[width=0.47\textwidth]{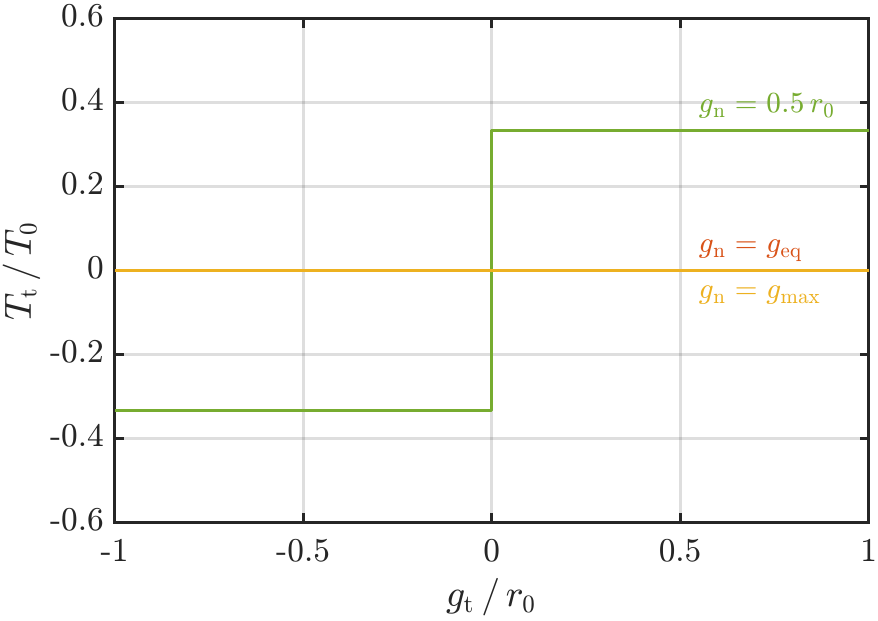}
		\label{f:TtlawAM:s0}
	}
	\vspace*{1ex}

	\subfigure[Sliding threshold for varying $\gn$; $\scut = 0.5$.]{
		\includegraphics[width=0.47\textwidth]{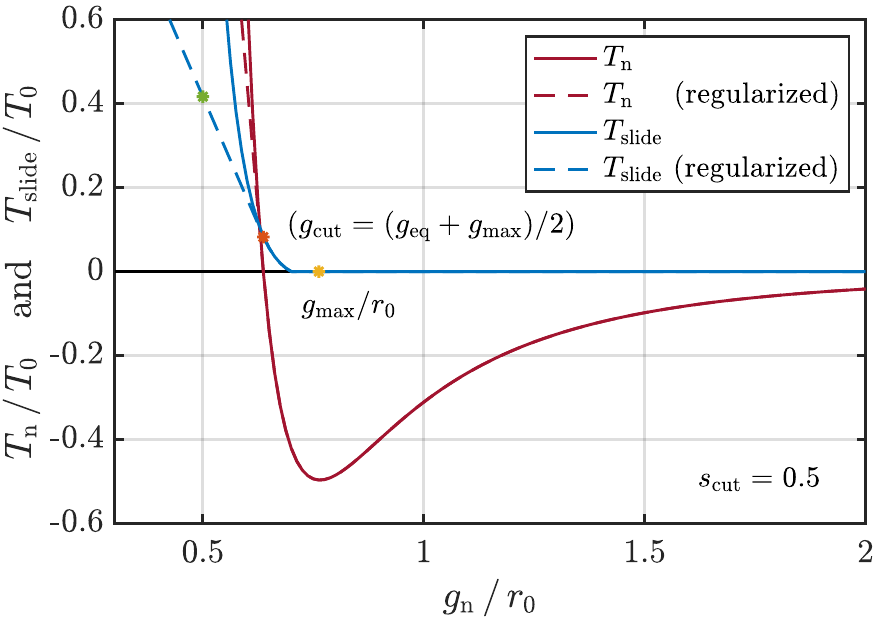}
		\label{f:TslideAM:s05}
	}\hspace*{1ex}
	\subfigure[Friction law for different but fixed $\gn$; $\scut = 0.5$.]{
		\includegraphics[width=0.47\textwidth]{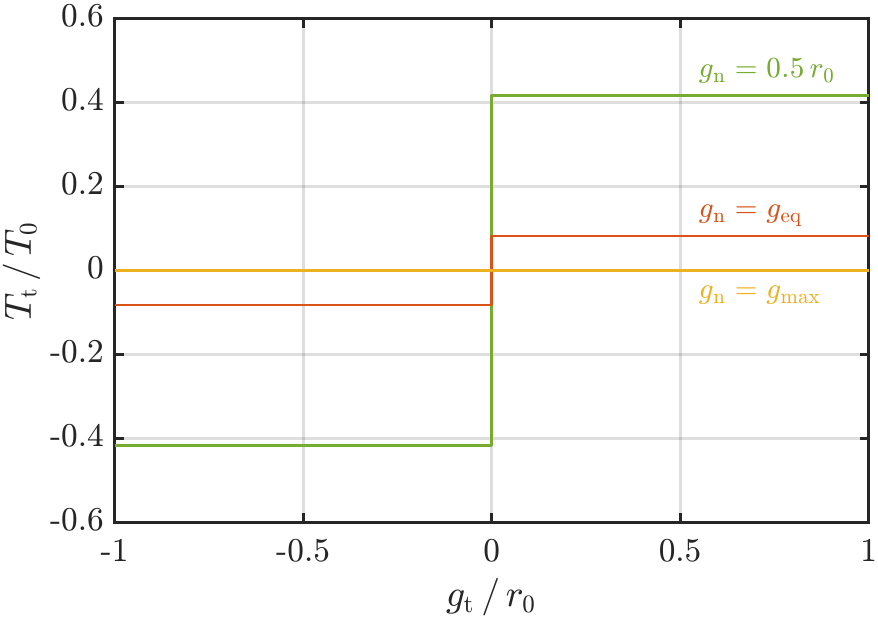}
	}
	\vspace*{1ex}

	\subfigure[Sliding threshold for varying $\gn$; $\scut = 1$.]{
		\includegraphics[width=0.47\textwidth]{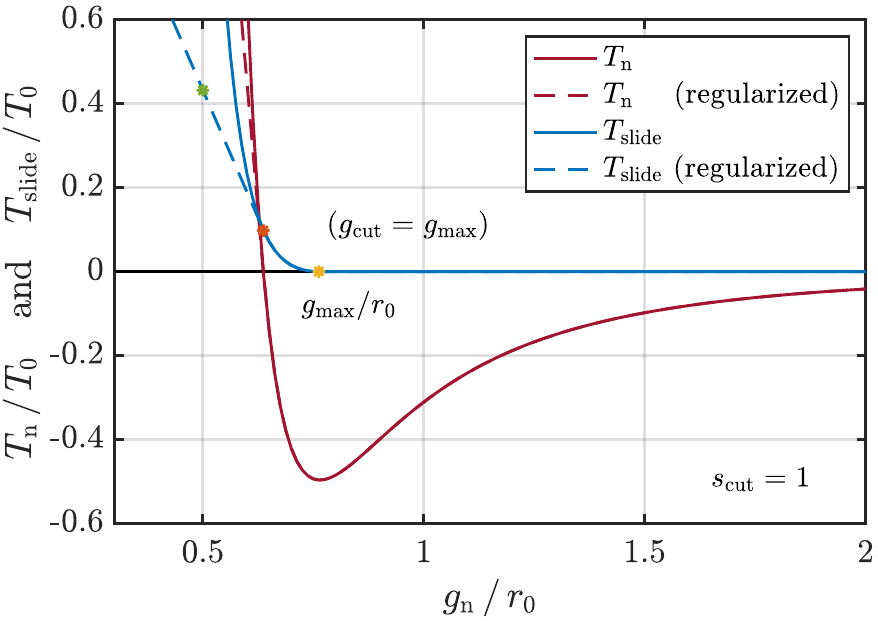}
		\label{f:TslideAM:s1}
	}\hspace*{1ex}
	\subfigure[Friction law for different but fixed $\gn$; $\scut = 1$.]{
		\includegraphics[width=0.47\textwidth]{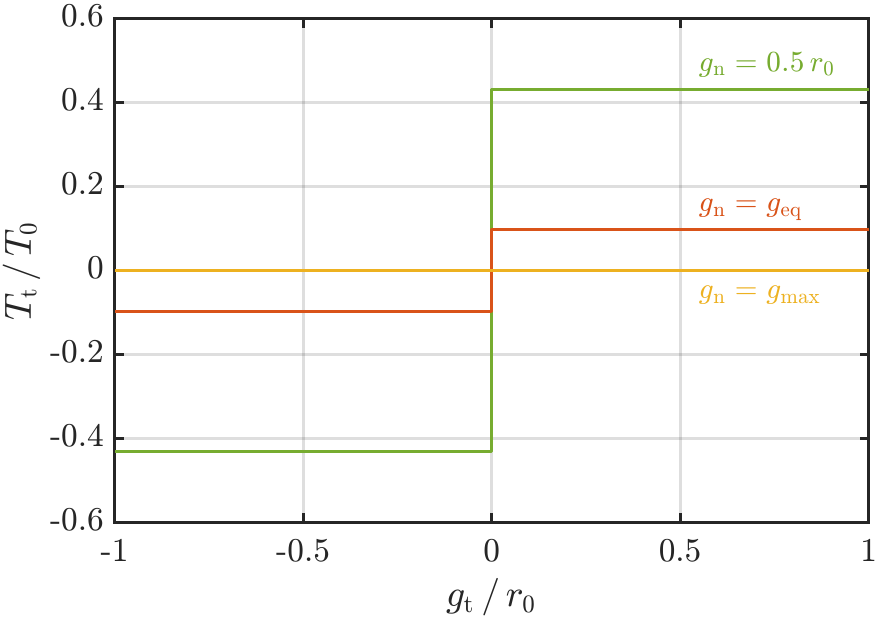}
		\label{f:TtlawAM:s1}
	}
	\caption{Model~\AM: Extended Amontons' law in \emph{local} form
		illustrated for $\mu_\AMe = 0.2$, $\Jcl \equiv 1$, and three different
		values of~$\scut$; $T_0 = \AH \,/\, (2\pi r_0^3)$; the colored
		asterisks mark the normal gap values $0.5\,r_0$~(green),
		$\gequ$~(where $\Tn = 0$, orange), and $\gmax$~(yellow).}
	\label{f:modelAM}
\end{figure}
Let us first shift the traction~$\Tn$ in \cref{e:Tn} by a value smaller than or
equal to the maximum adhesive traction, $\Tnmax$; see also
\cref{e:gmaxTnmax,f:Tn}. To this end, we introduce a distance, $\gcut$, which
lies somewhere between the equilibrium distance,~$\gequ$, and the
location~$\gmax$ of~$\Tnmax$:
\begin{equation}
	\gcut = \scut \, \gmax + (1 - \scut) \, \gequ, \qquad \scut \in [0,1].
	\label{e:gcutAM}
\end{equation}
As can be seen in \cref{f:TslideAM:s0,f:TslideAM:s1}, in this range
$\Tn(\gcut)$ is smaller than or equal to zero. We then consider a sliding
threshold that is proportional to the shifted curve $\Tn(\gn) + |\Tn(\gcut)|$:
\begin{equation}
	\Ttmax(\gn) = \begin{cases}
		\displaystyle
		\frac{\mu_\AMe}{\Jcl} \, \big[\Tn(\gn) - \Tn(\gcut)\big],
			& \gn < \gcut, \\[0.2ex]
		0,
			& \gn \ge \gcut.
	\end{cases} \label{e:TslideAM}
\end{equation}
Note that in this case, the sliding threshold,~$\Ttmax$, directly depends on
the normal traction,~$\Tn$, which is defined in the \emph{reference}
configuration (\cref{s:model:CGCM}).

\Cref{f:modelAM} illustrates model~\AM\ for three different values of the
parameter~$\scut$. The left-hand side of the figure shows the dependence of the
sliding traction,~$\Ttmax$, on the normal gap,~$\gn$; dashed lines indicate a
regularized version according to \cref{a:regularize}. If $\scut = 0$
(\cref{f:TslideAM:s0,f:TtlawAM:s0}), tangential sliding occurs only for
positive, i.e.~compressive, normal tractions. This corresponds to classical
Coulomb-Amontons friction (see \cref{e:Coulomb:tract}) for non-adhesive
contact. It is further used in many cohesive zone models (see
\cref{s:adhfrict:others}) to include frictional sliding. If $\scut > 0$, a
tangential sliding resistance is present even for tensile normal tractions.
Note that the curve for $\Ttmax$ is smooth ($C^1$-continuous) only for $\scut =
1$ (\cref{f:TslideAM:s1}); otherwise, a kink occurs at $\gn = \gcut$, which
requires special treatment in a computational implementation \cite{mergelPhD}.
For the particular case $\scut = 0$, this kink is exactly located at the
equilibrium position, $\gequ$.
\begin{figure}[ht]
	\centering
	\subfigure[Model~\AM\ with $\gcut = \gequ$.]{
		\includegraphics[width=0.47\textwidth]{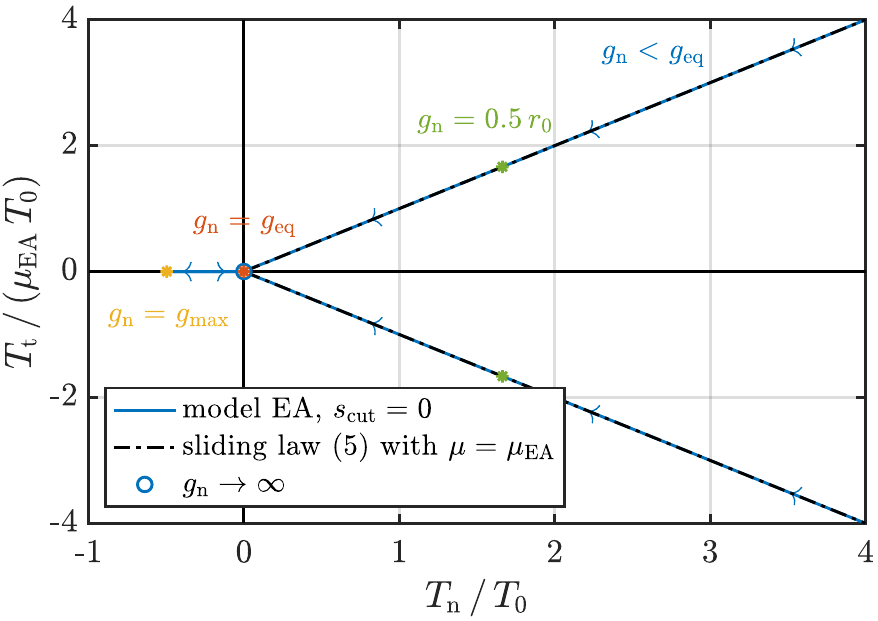}
	}\hspace*{1ex}
	\subfigure[Model~\AM\ with $\gcut = (\gequ+\gmax)/2$.]{
		\includegraphics[width=0.47\textwidth]{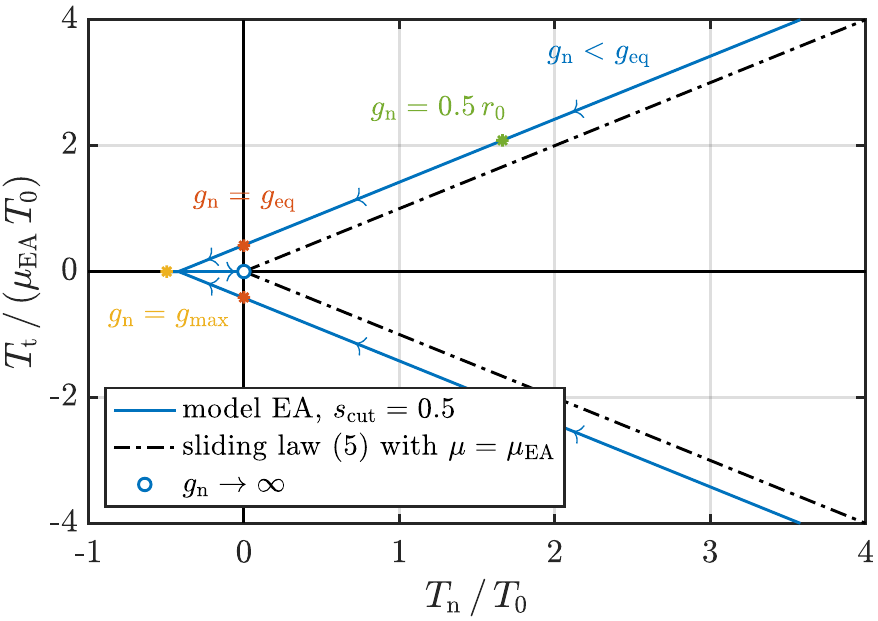}
	}
	\subfigure[Model~\AM\ with $\gcut = \gmax$.]{
		\includegraphics[width=0.47\textwidth]{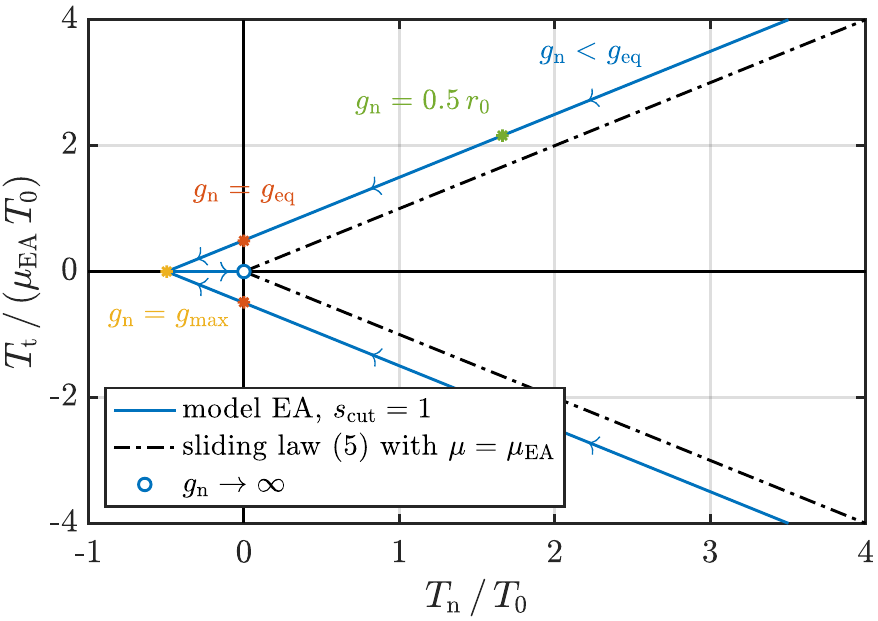}
	}\hspace*{1ex}
	\subfigure[Model~\CA.]{
		\includegraphics[width=0.47\textwidth]{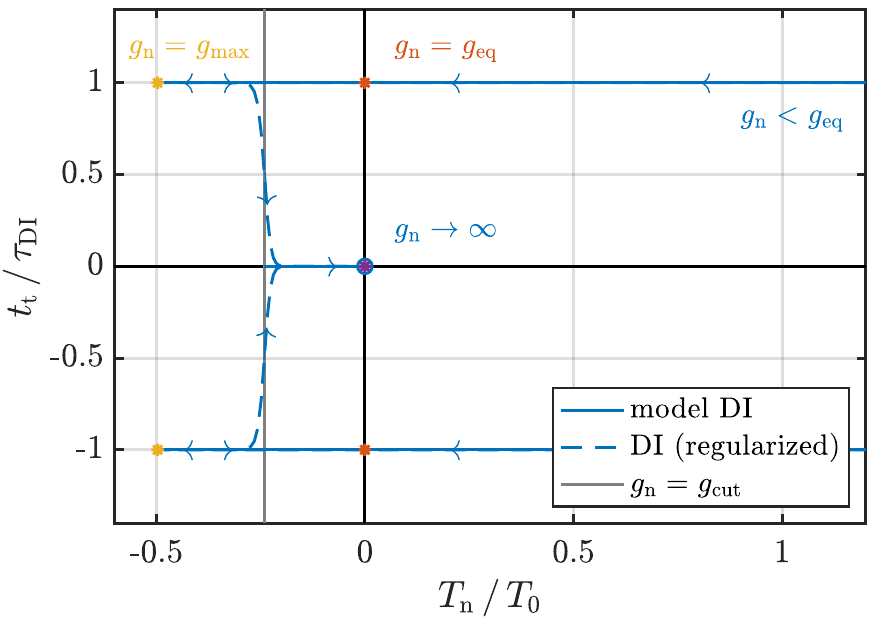}
		\label{f:TnTt:CA}
	}
	\caption{Tangential traction during sliding vs.~normal traction for
		(a) -- (c) model~\AM\ with $\Jcl \equiv 1$ and (d)~model~\CA;
		$T_0 = \AH \,/\, (2\pi r_0^3)$; the blue arrows indicate the sequence
		from $\gn = 0$ to $\gn \to \infty$; for the colored asterisks see
		\cref{f:modelCA,f:modelAM}, respectively.}
	\label{f:TnTt}
\end{figure}

\cref{f:TnTt} depicts the (directed) tangential traction during sliding versus
the normal traction, either for model~\AM\ and different values of~$\scut$, or
for model~\CA. As shown, for model~\AM\ this relation resembles a shifted
version of the cone describing the classical friction law
\labelcref{e:Coulomb:tract}. In fact, up to $\gn \le \gcut$, it corresponds to
the Mohr-Coulomb model mentioned in \cref{s:adhfrict:others}. In model~\CA, the
tangential traction, $\ttsmall$, is not a function of the normal traction,
$\tn$, anymore.


\subsection{Continuum mechanical equations}

For completeness this section outlines the governing equations for
adhesive and frictional contact of two bodies. Under quasi-static conditions
(for which inertial forces are neglected) the following equilibrium equations
must be satisfied at each point~$\xk$ within the two bodies $\sB_k$ $(k =
1,\,2)$,
\begin{equation}
	\divv\,\bsig_k + \bff_k = \bzero. \label{e:equilibrium}
\end{equation}
Here, $\bsig_k$ is the Cauchy stress tensor, and $\bff_k$ denotes the vector of
distributed volumetric forces. The bodies must further satisfy the contact
condition
\begin{equation}
	\bsig_k \, \bn_k = \btnk - \bttk \label{e:contC}
\end{equation}
on the contact surface~$\partial_{\mrc} \sB_k$, as well as conditions at the
Dirichlet boundaries~$\partial_{\bu}\sB_k$ (where displacements~$\bar{\bu}_k$
are prescribed)
\begin{equation}
	\bx_k - \bX_k = \bar{\bu}_k,
\end{equation}
and at the Neumann boundaries~$\partial_{\bt}\sB_k$ (where surface
tractions~$\bar{\bt}_k$ are applied),
\begin{equation}
	\bsig_k \, \bn_k = \bar{\bt}_k. \label{e:NeumannBC}
\end{equation}
In \cref{e:contC,e:NeumannBC}, the unit vector~$\bn_k$ denotes the current
surface normal at point~$\xk$. \Cref{e:contC} further contains the normal and
tangential tractions defined through our contact models; note that here the
sign convention of Laursen~\cite{laursen02} is used.
\Crefrange{e:equilibrium}{e:NeumannBC} represent the strong form of the general
contact and boundary value problem, expressed in the current configurations of
the bodies. Since the analytical solution of these equations is possible only
for very special cases (and mainly restricted to small deformations), for
general conditions involving arbitrary geometries and large deformations, a
computational solution technique is required. For this reason, \cite{mergelPhD,
mergel18comput} provide a computational implementation of our models into a
nonlinear finite element formulation, and also discuss the algorithmic
treatment of friction under large deformations.


\subsection{General comments, validity, and restrictions}
\label{s:model:validity}

Like any other model, our models have limitations. We address those in the
following.



In \cref{e:ttorig} we assume that the static friction threshold coincides with
the resistance for kinetic friction. As mentioned at the beginning of
\cref{s:model:frict}, this assumption agrees well with experimental
observations for bio-adhesive systems and rough elastomers. Nevertheless, these
observations refer to the \emph{globally} measured force resultant instead of
the \emph{local} traction at the contact zone. It remains to be discussed
further whether static and kinetic friction coincide because of the material
itself, or whether this is caused by a split of the effective contact area into
a large number of small areas. The influence of such a split on both static
friction and stick-slip motion is addressed in~\cite{varenberg07, varenberg09,
lorenz12}. Regarding bio-inspired adhesives, it may be arguable whether the
assumption of equal static and kinetic friction is valid for all kinds of
materials. See, for instance, the experimental results~\cite{varenberg07}
obtained for a microstructured polyvinylsiloxane (PVS) surface sliding on
glass. If required, our model could be extended to account for differing
parameters for static and kinetic friction as well.



In our model we consider dry (i.e., non-lubricated) adhesion and friction. We
hence omit the influence of any secretion that may cover the adhesive device,
as observed for many beetles and other insects. In fact, experimental studies
on insect pads show that the depletion of such secretion (caused e.g.~by
sliding or by repeating detachment) affects both their frictional
resistance~\cite{drechsler06, bullock08} and (although less strongly) their
adhesion~\cite{labonte15}. Nevertheless, if the amount of sliding is
sufficiently small, one can assume that the frictional resistance does not
change considerably.



As shown in \cref{s:intro,s:exp}, there exist several applications for which
van der Waals forces may affect the macroscopic behavior even at larger scales
(micrometers or millimeters). For increasing length scales, a direct
computational implementation of adhesion models based on the Lennard-Jones
potential, such as the CGCM model (\cref{s:model:CGCM}) and friction model~\AM,
can become very inefficient, because they require nanoscale finite element
resolution. For this reason, it would be very promising to develop an effective
adhesion model that is regularized by the compliance of the surrounding
material. One approach in this direction has been developed in the context of
adhesive joints under small deformations \cite{schmidt10}. Another possibility,
which is pursued in~\cite{mergelPhD, mergel18comput} and also here, is to
calibrate the parameters in \cref{s:model:CGCM} (like $\AH$, $\Tnmax$, or
$W_\adh$) such that they match with experimental data. The
curve~\labelcref{e:Tn} is then regularized by an automatic increase of the
length parameter~$r_0$. As a consequence, the model does not require nanoscale
resolution anymore. When adjusting these parameters it is important to
distinguish between the nominal (or apparent) contact area and the true contact
area due to very small asperities, which interact at the contact surface; see
also the discussion in \cref{s:originfrict}. For the CGCM of
\cref{s:model:CGCM} the normal contact stress is integrated over an apparent,
nominally flat contact area. Thus, simply inserting the material constant~$\AH$
for the considered pair of materials would overestimate the real strength of
adhesion by several orders of magnitude. This can be overcome by directly
inserting effectively measured values as described before. Note that in
general, the ratio between the true and the nominal contact areas may depend on
the contact pressure, as discussed in Sect.~5.2.4 of~\cite{wriggers06} from a
computational point of view.


\section{Qualitative comparison between model and experiment} \label{s:example}

\begin{figure}[ht]
	\centering
	\subfigure[Problem setup.]{
		\includegraphics[width=0.48\textwidth]{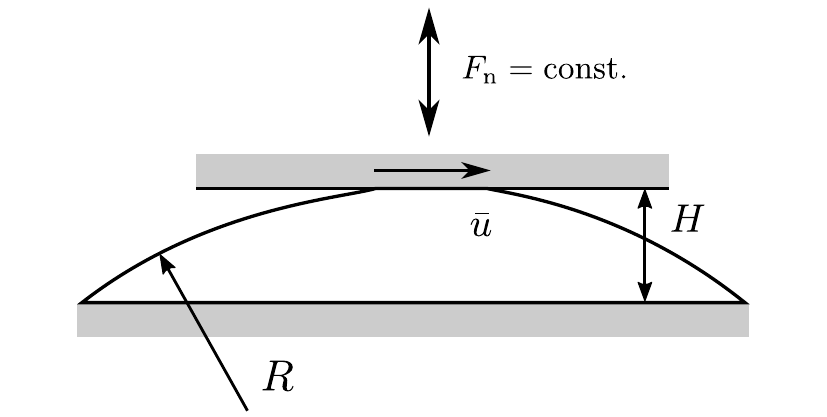}
		\label{f:cap:setup}
	}
	\subfigure[Sliding under zero normal load.]{
		\unitlength\textwidth
		\begin{picture}(0.48,0.1)
		\put(0,0.0615){\includegraphics[width=0.48\textwidth]{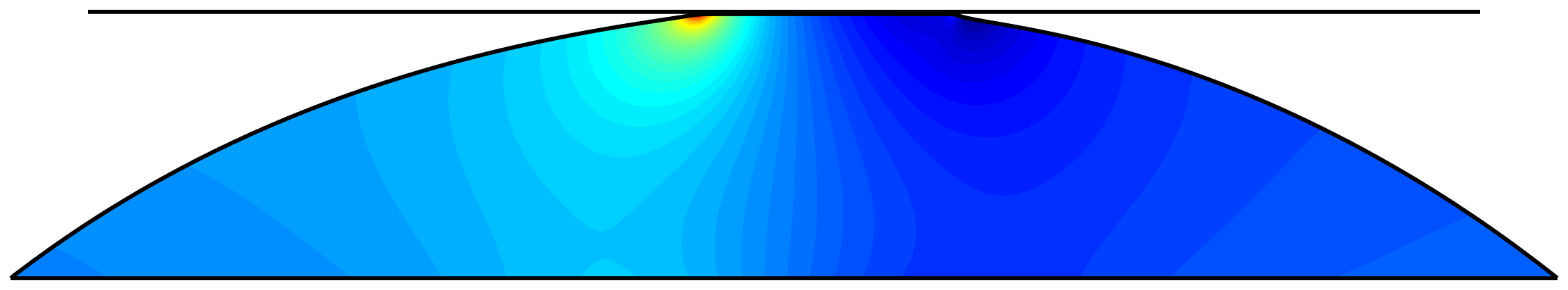}}
		\put(0.01,0){\includegraphics[width=0.46\textwidth]{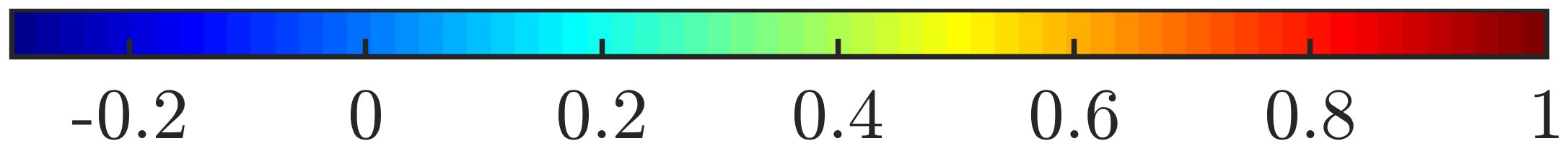}}
		\end{picture}
		\label{f:cap:xn}
	}
	\caption{Adhesive friction of a soft cap: (a)~Setup; (b)~stress
		distribution during sliding for zero preload and model~\CA\ ($\gcut
		= \gmax$, $\mu_\CAe = 1$); the colors show the first invariant of the
		Cauchy stress in units of~$E$.}
\end{figure}

We now apply the proposed two models to study adhesive friction between a soft
cylindrical cap and a rigid substrate, see \cref{f:cap:setup}. To this end, we
use the finite element implementation derived in~\cite{mergelPhD,
mergel18comput}.

We consider this example as a preliminary study to investigate the qualitative
behavior of our models, and to show that it agrees with the experimental
behavior presented in \cref{s:exp}. For length scale reasons related to the
computational adhesion model~\cite{sauer09cmame} (addressed in detail in
\cref{s:model:validity} already), a quantitative comparison is not possible at
this stage. Therefore, the cap in the example is 2D and smaller than that in
\cref{s:exp}. Also, slightly different material parameters (\cref{t:cap:para})
and smaller friction values are used. A 2D plane strain, nonlinear finite
element formulation based on a Neo-Hooke material model is used to simulate the
example. For further details see~\cite{mergelPhD}. All results shown here are
normalized by Young's modulus, $E$, an arbitrary out-of-plane thickness (or
width), $W$, and a unit length, $L_0$ (see \cref{t:cap:para}).
\begin{table}[h]
	\centering
	\begin{tabular}
		{c@{\qquad}c@{\qquad}c@{\qquad}c@{\qquad}c@{\qquad}c@{\qquad}c}
		\hline \noalign{\smallskip}
		$E$ & $\nu$ & $\Tnmax$ & $W_\adh$ & $R$	& H & $L_0$ \\
		\noalign{\smallskip} \hline \noalign{\smallskip}
		$2\,\mrM\mrP\mra$ & $0.4$ & $0.33\,\mrM\mrP\mra$
		& $0.027\,\mrJ/\mrm^2$ & $47.1\,L_0$ & $10\,L_0$ & $1\,\mu\mrm$ \\
		\noalign{\smallskip} \hline
	\end{tabular}
	\caption{Adhesive friction of a soft cap: Parameters.}
	\label{t:cap:para}
\end{table}

As illustrated in \cref{f:cap:setup}, we first apply a fixed normal force to
the rigid substrate, and then slide it horizontally while keeping the lower
boundary of the cap fixed. As a special case we investigate the sliding
behavior also under zero load, for which the attractive and repulsive stresses
in the contact area equilibrate each other. The finite element mesh of the cap
consists of 42,300 Q1N2.1 elements \cite{corbett14}. \Cref{f:cap:xn} shows the
stress distribution of the cap during sliding under zero load.

Although keeping the normal load constant during sliding is not strictly
identical to the boundary conditions in our experiment, in the latter the
normal load is expected to remain reasonably close to its initial value
predicted from the JKR theory (\cref{f:exp:FnJKR}): Since the end of the double
cantilever (see \cref{s:exp}) is kept vertically fixed, the vertical position
of the glass plate, and thus the normal load, can change only due to a vertical
dilatancy of the elastomer sphere caused by horizontal shear. According to
Scheibert et al.~\cite{scheibert09} (see the introduction and references
therein), however, in our case of a rigid body (glass plate) in contact with an
incompressible half-space (thick elastomer with a Poisson's ratio of
approx.~$0.5$), the coupling between the normal displacement and any induced
tangential stresses is expected to be negligible. For a qualitative comparison
as it is performed here, it is expected that the slight differences in the
boundary conditions do not affect our general observations and conclusions.

\begin{figure}[p]
	\centering
	\subfigure[Friction force under compression, $\Fn^* = -0.288$.]{
		\includegraphics[width=0.47\textwidth]{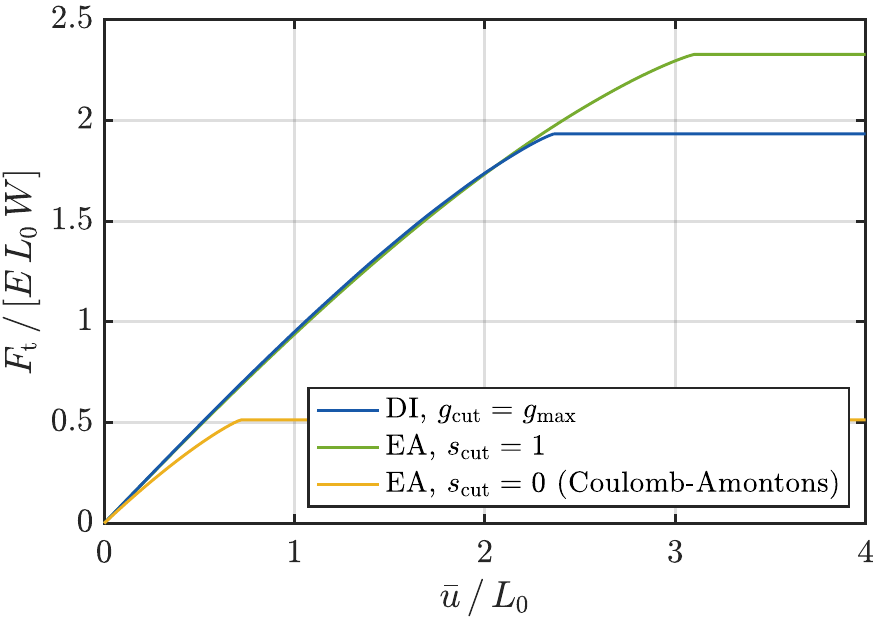}
	}\hspace*{1ex}
	\subfigure[Contact area under compression, $\Fn^* = -0.288$.]{
		\includegraphics[width=0.47\textwidth]{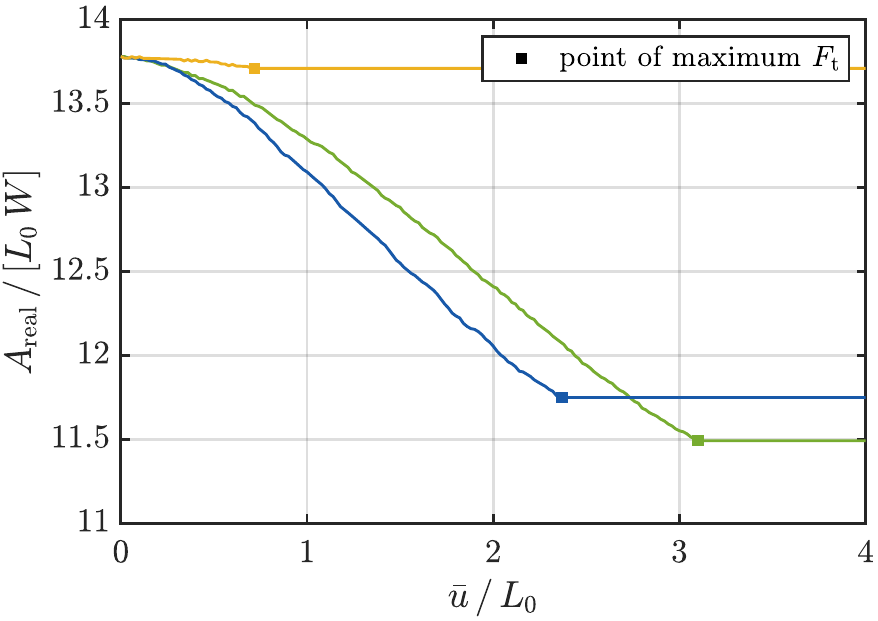}
	}
	\subfigure[Friction force under zero load, $\Fn^* = 0$.]{
		\includegraphics[width=0.47\textwidth]{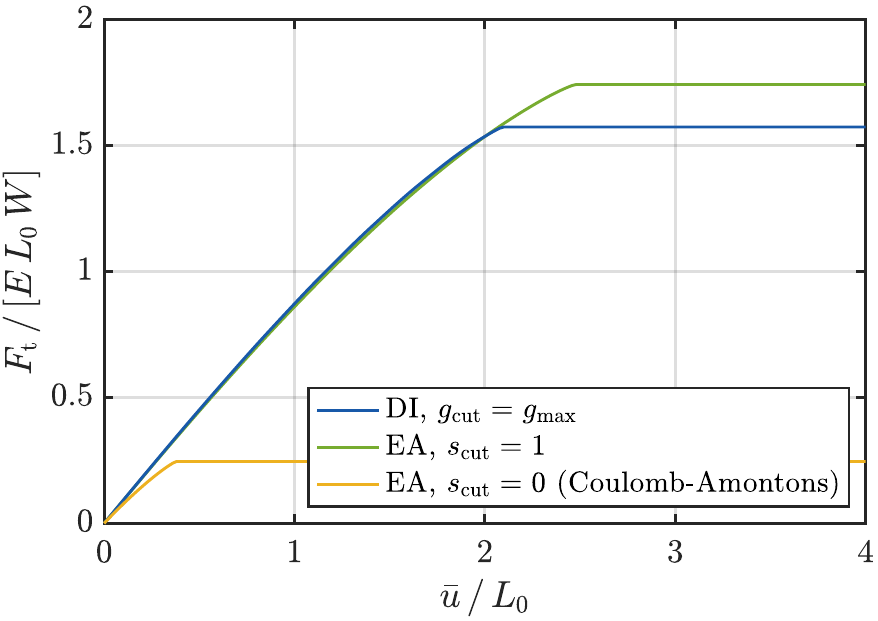}
	}\hspace*{1ex}
	\subfigure[Contact area under zero load, $\Fn^* = 0$.]{
		\includegraphics[width=0.47\textwidth]{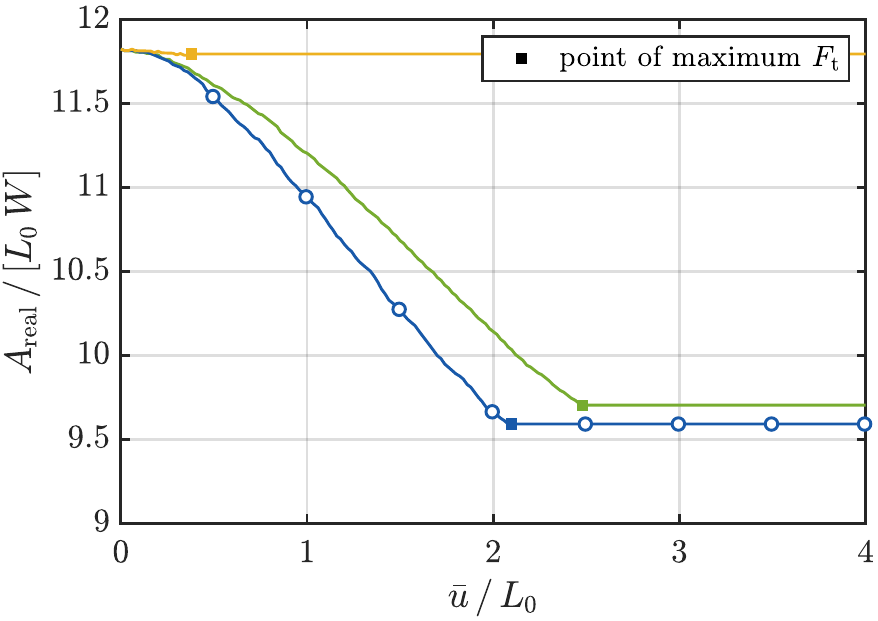}
		\label{f:cap:Ac0}
	}
	\subfigure[Friction force under tension, $\Fn^* = 0.288$.]{
		\includegraphics[width=0.47\textwidth]{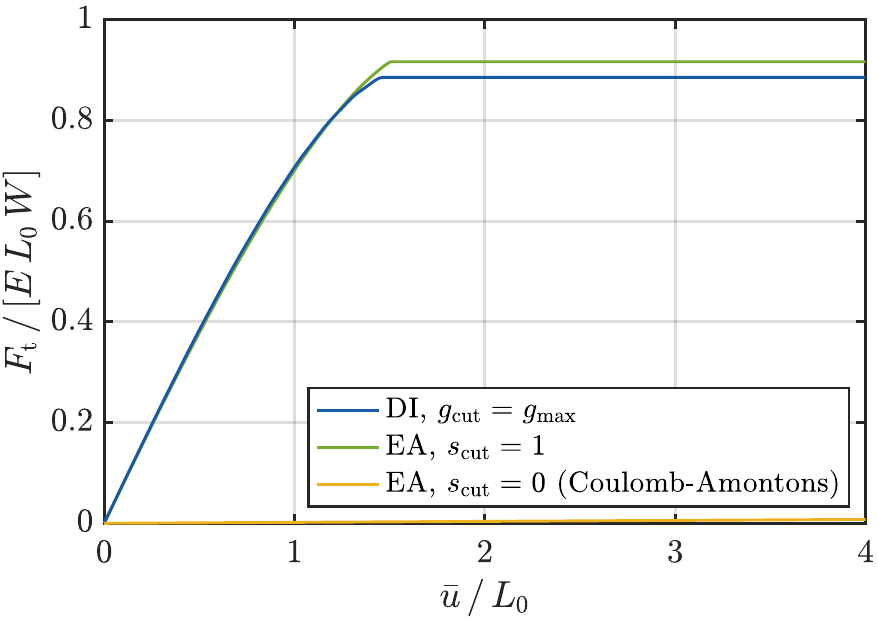}
	}\hspace*{1ex}
	\subfigure[Contact area under tension, $\Fn^* = 0.288$.]{
		\includegraphics[width=0.47\textwidth]{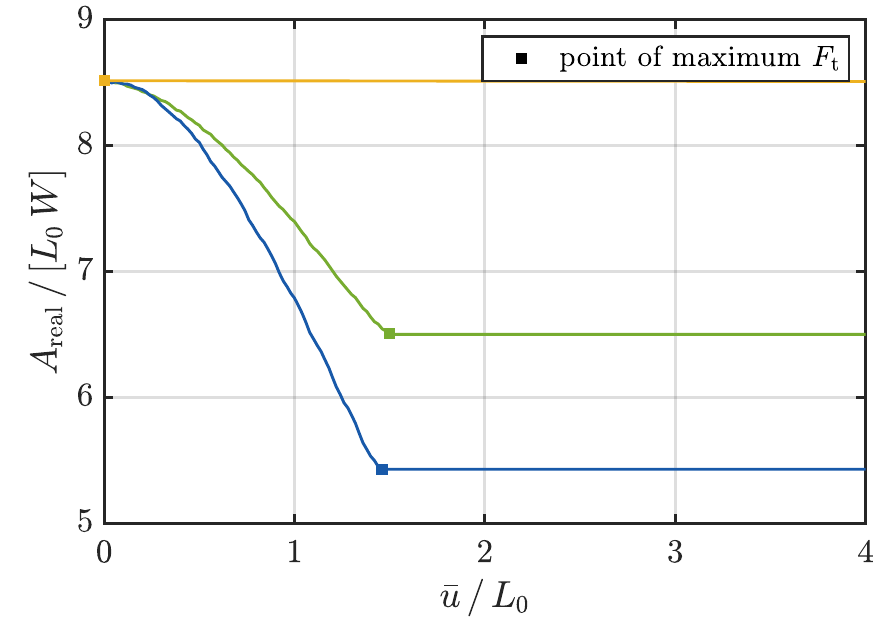}
	}
	\caption{Adhesive friction of a soft cap: Friction force and contact area
		for sliding under a constant preload ($\mu_\AMe = \mu_\CAe = 1$); the
		squares indicate the onset of full sliding; the dots mark the
		configurations of \cref{f:cap:Aczoom}; $\Fn^* = \Fn \,/\,
		\big[E\,L_0\,W\big]$ denotes the normalized preload.}
	\label{f:cap:PxAndAc}
\end{figure}
\Cref{f:cap:PxAndAc} shows both the friction force and the contact area for
different loads. Here we use models~\CA\ with $\gcut = \gmax$, \AM\ with $\scut
= 1$, and \AM\ with $\scut = 0$ (Coulomb-Amontons friction, see
also~\cref{e:Coulomb:tract}). Before full sliding, the qualitative behavior of
all models is close to our experimental results in
\cref{f:exp:FtTime,f:exp:AcTime}. For classical Coulomb-Amontons friction,
however, the sliding force is considerably lower than for the other models,
because the compressed area is much smaller than the total contact area. When
the entire contact surface is sliding, both the force and the area remain
constant for the numerical models. In contrast, in the experiments, the
tangential force drops over a finite time scale before entering a rather steady
sliding regime. These differences are most likely caused by the viscous
behavior of polydimethylsiloxane. It thus makes sense to later consider a
viscoelastic material model.

\begin{figure}[h]
	\centering
	\includegraphics[width=0.49\textwidth]{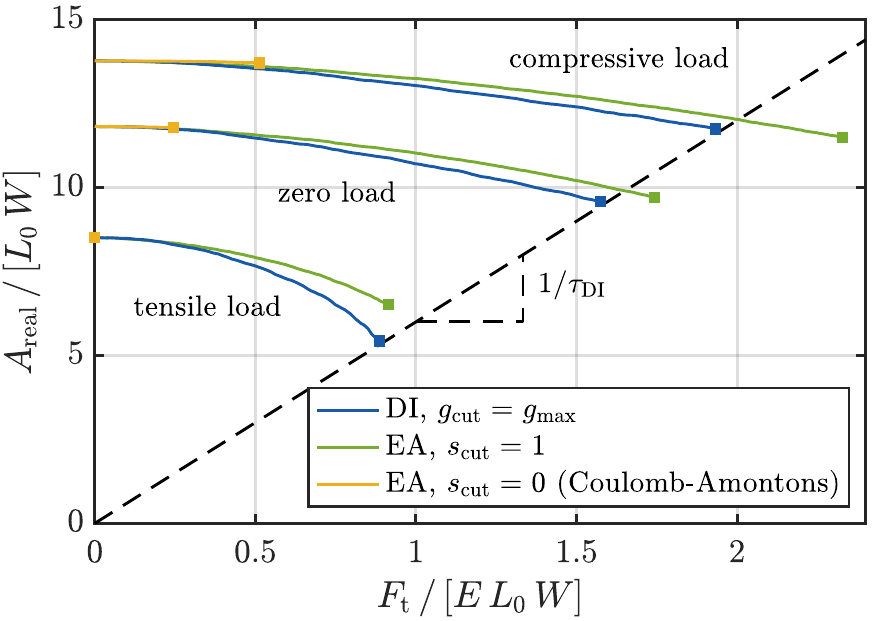}
	\caption{Adhesive friction of a soft cap: Contact area as a function of the
		tangential force for the onset of sliding under different preloads
		($\mu_\AMe = \mu_\CAe = 1$); $\tau_\CAe = \mu_\CAe \, \Tnmax$; see also
		\cref{f:exp:AcFt}.}
	\label{f:cap:PxOverAc}
\end{figure}

\Cref{f:cap:PxOverAc} (like \cref{f:exp:AcFt}) shows the contact area versus
the tangential force up to the point of full sliding. The classical law by
Coulomb and Amontons (model \AM\ for $\scut = 0$) obviously fails to capture
the qualitative behavior observed in the experiments. In contrast, the
differences to the experimental results are smaller for model~\AM\ with $\scut
= 1$, but they are still considerable. Model \CA, on the other hand, agrees
very well with the experiments.

\begin{figure}[h]
	\unitlength\textwidth
	\begin{picture}(1,0.37)
		\put(0.32,0.34){\includegraphics[width=0.36\textwidth]{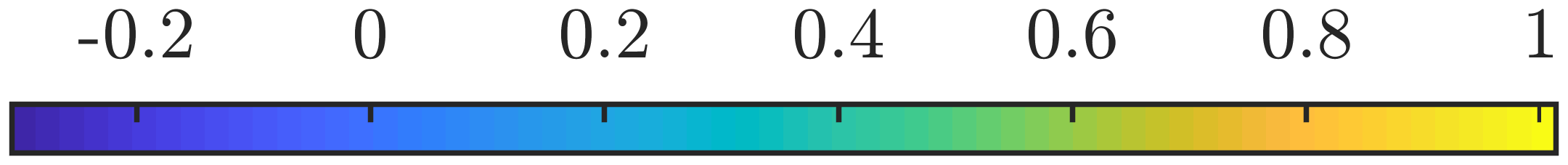}}
		\put(0,0.18){\includegraphics[width=0.226\textwidth]{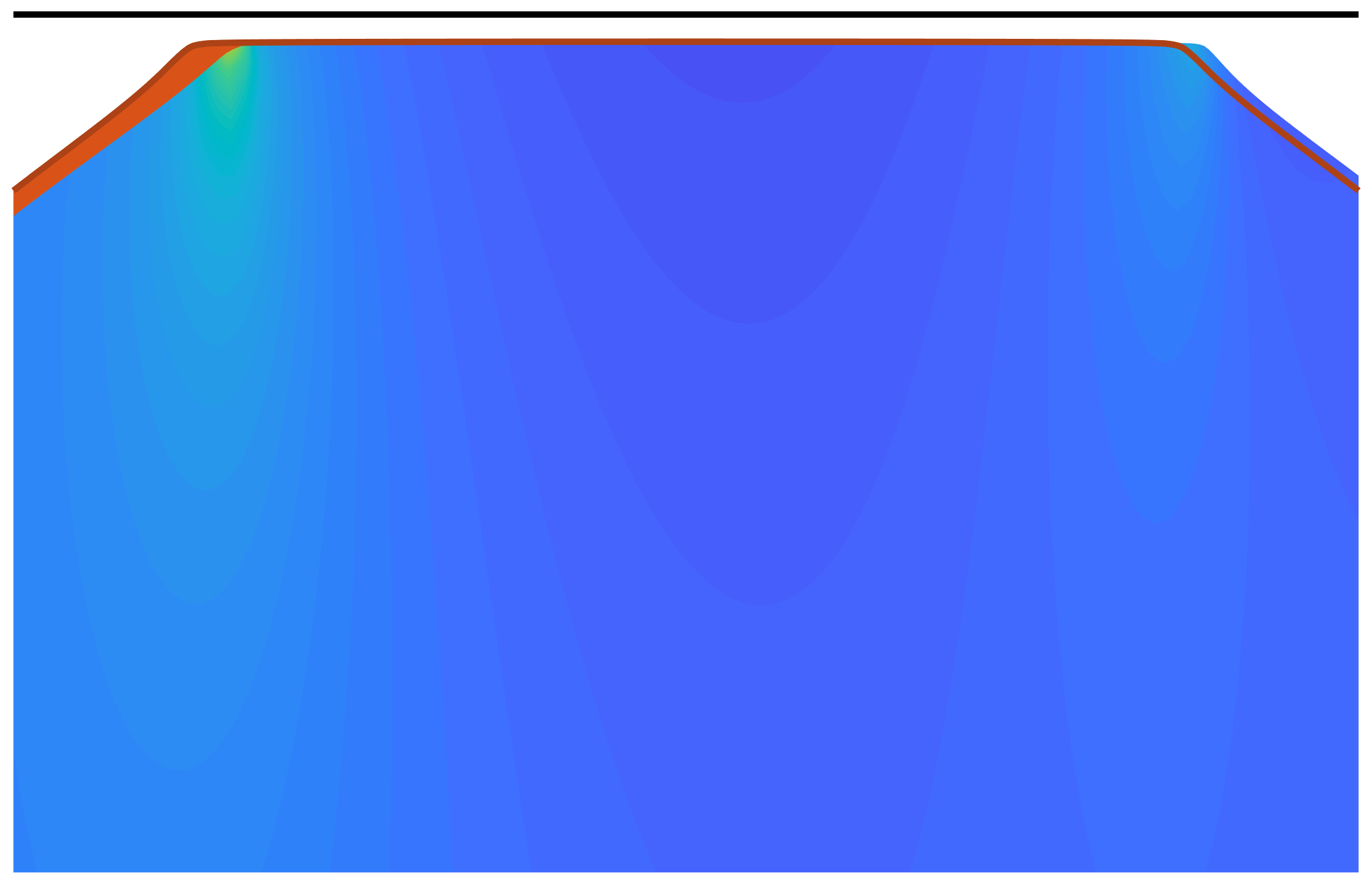}}
		\put(0.258,0.18){\includegraphics[width=0.226\textwidth]{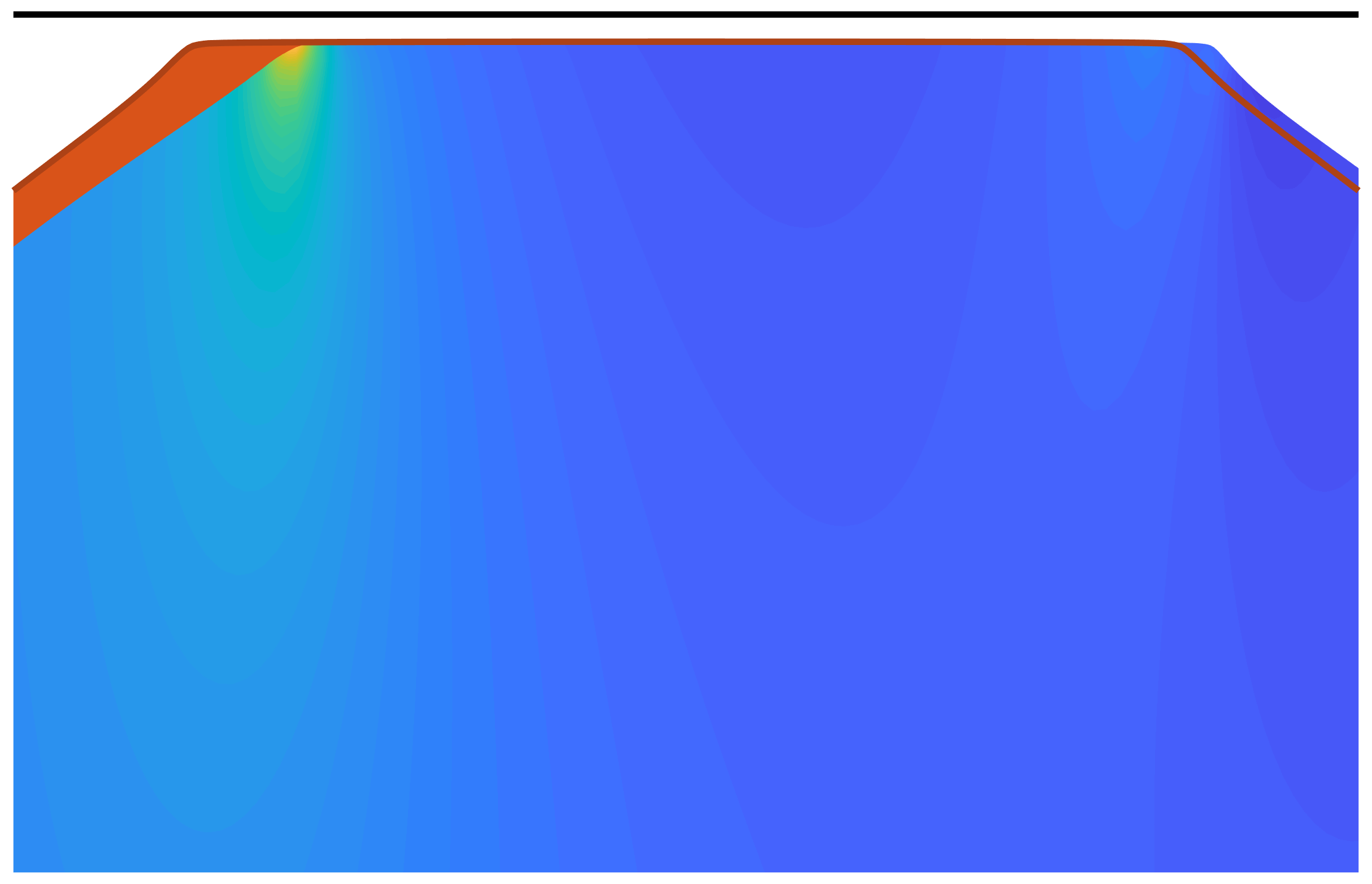}}
		\put(0.516,0.18){\includegraphics[width=0.226\textwidth]{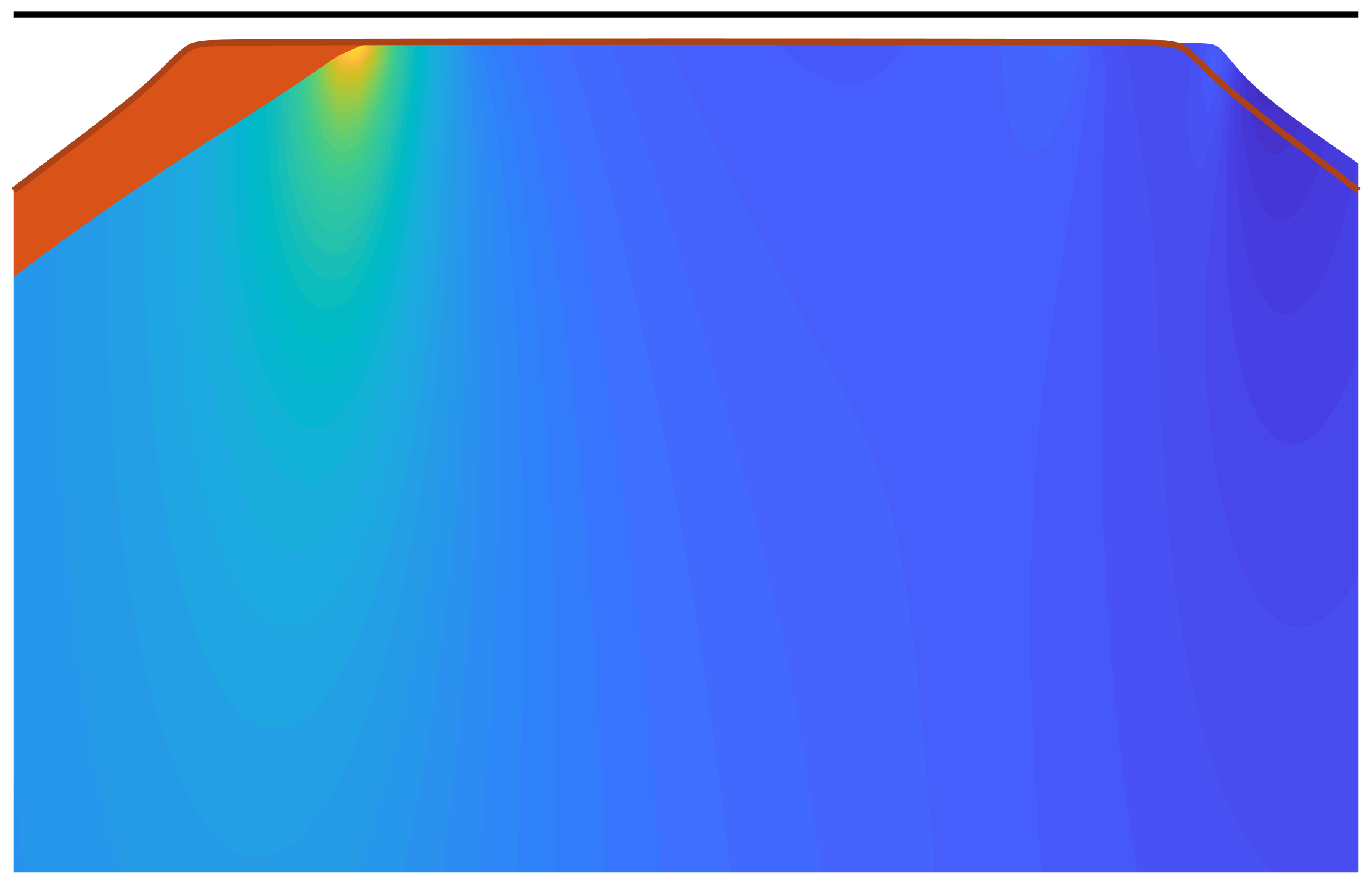}}
		\put(0.774,0.18){\includegraphics[width=0.226\textwidth]{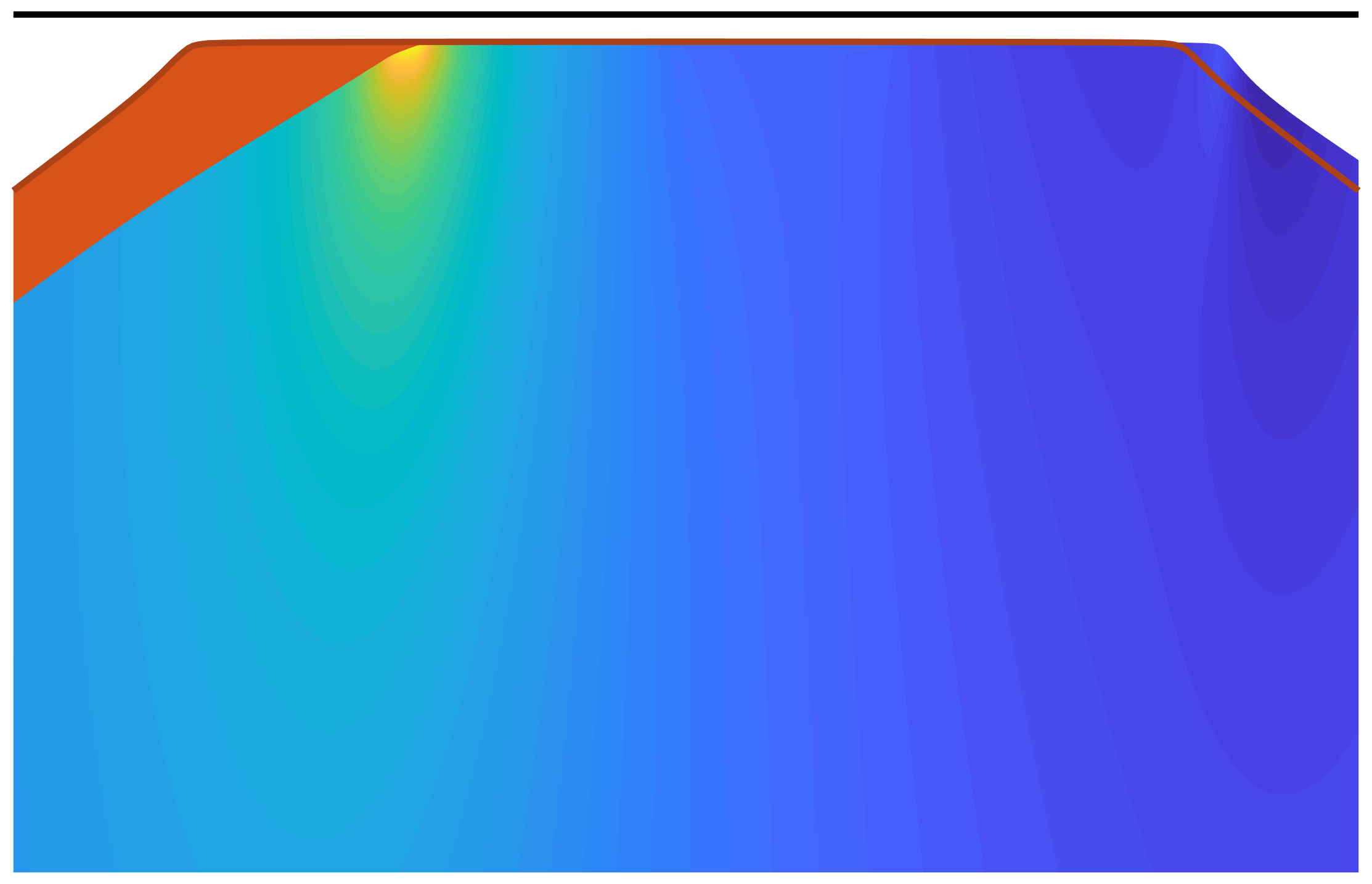}}
		\put(0,0){\includegraphics[width=0.226\textwidth]{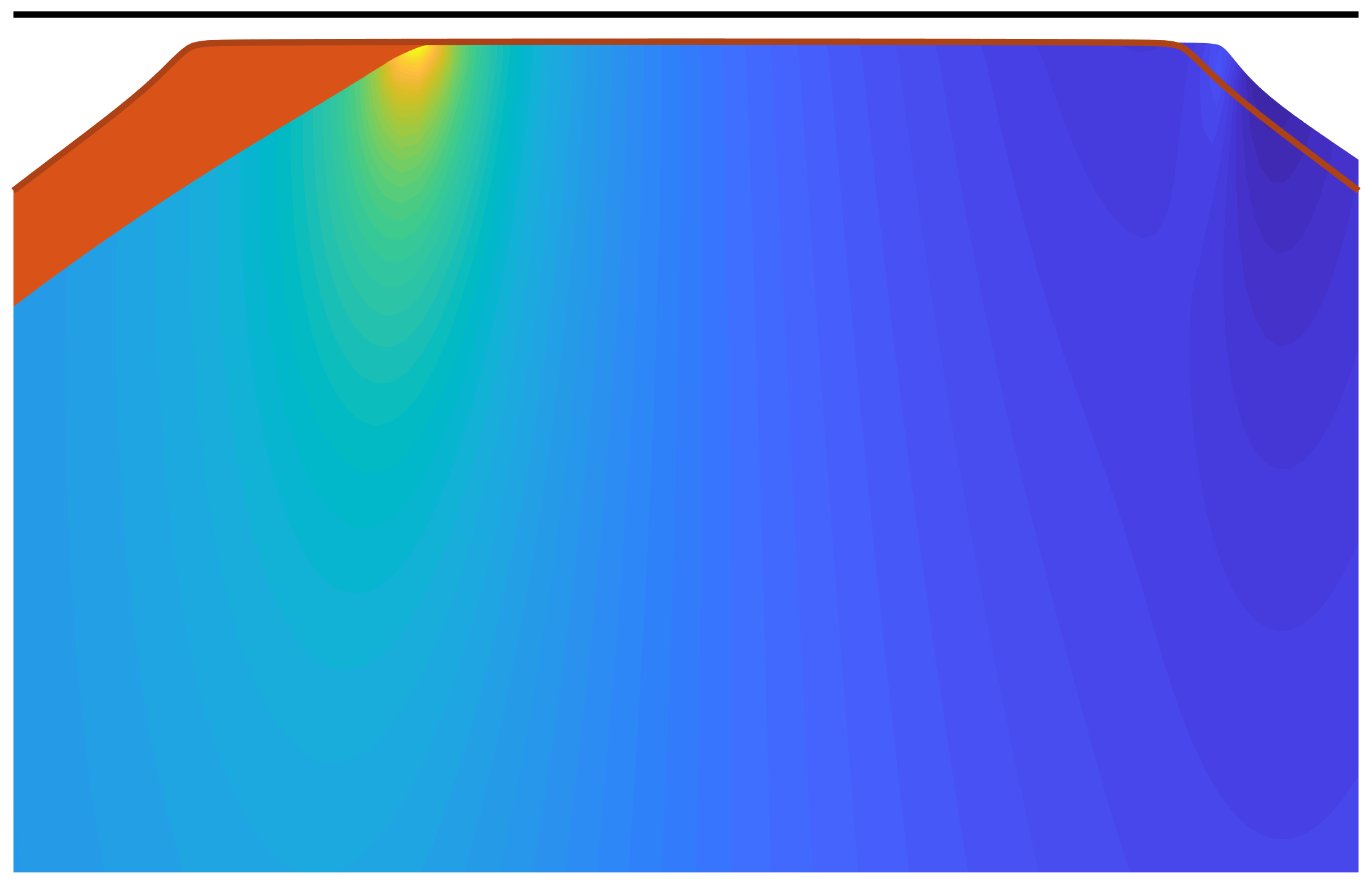}}
		\put(0.258,0){\includegraphics[width=0.226\textwidth]{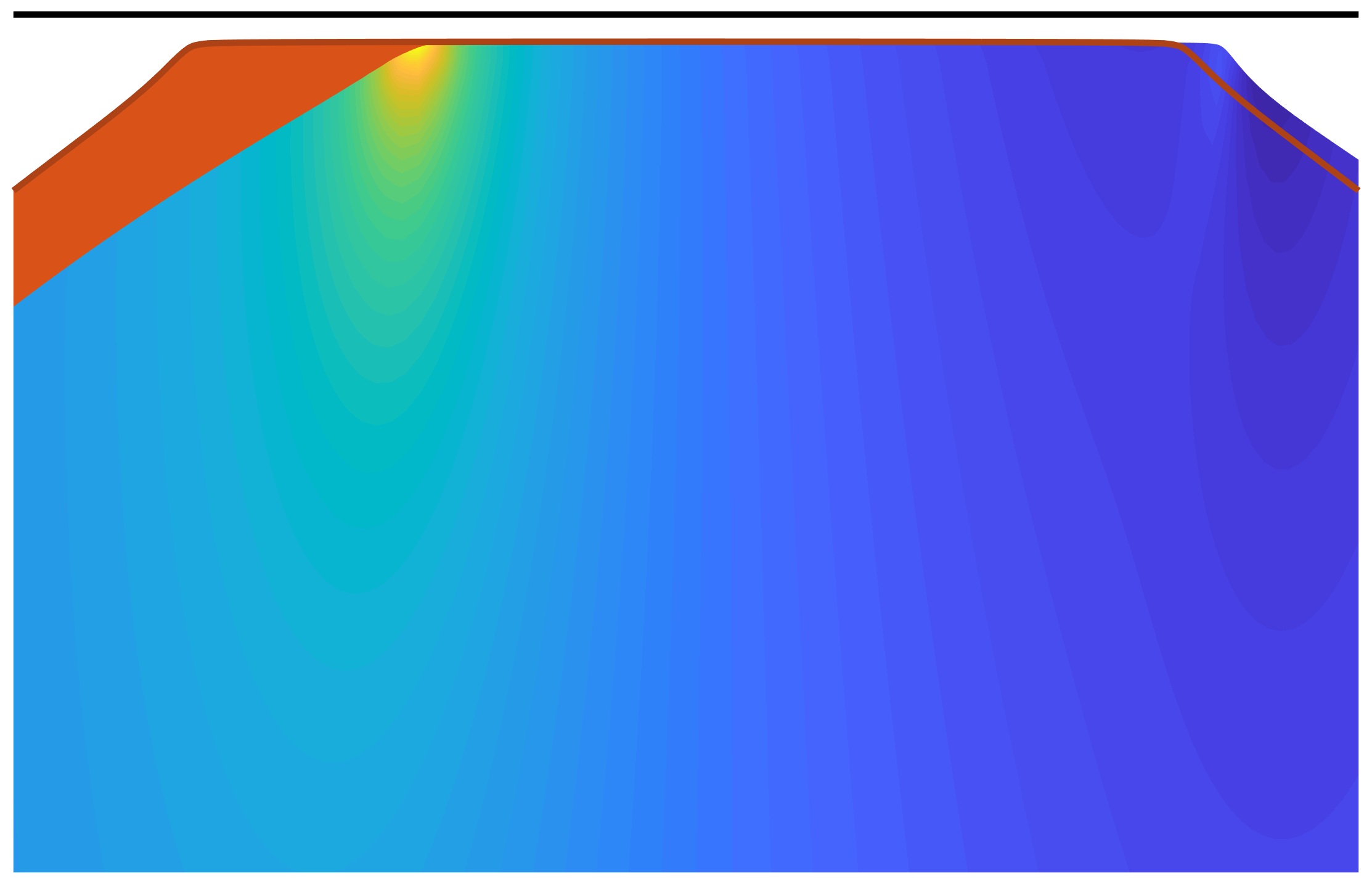}}
		\put(0.516,0){\includegraphics[width=0.226\textwidth]{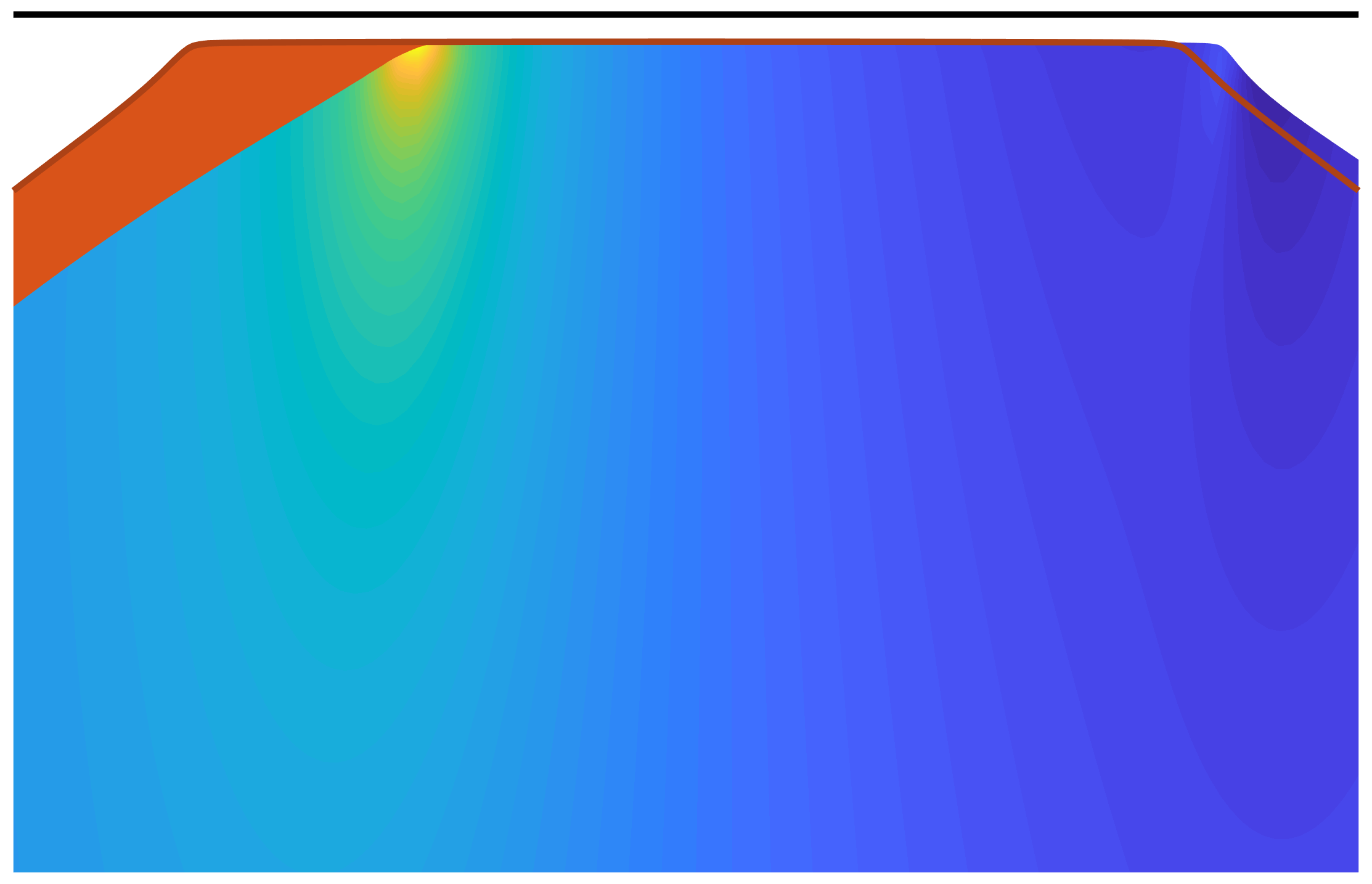}}
		\put(0.774,0){\includegraphics[width=0.226\textwidth]{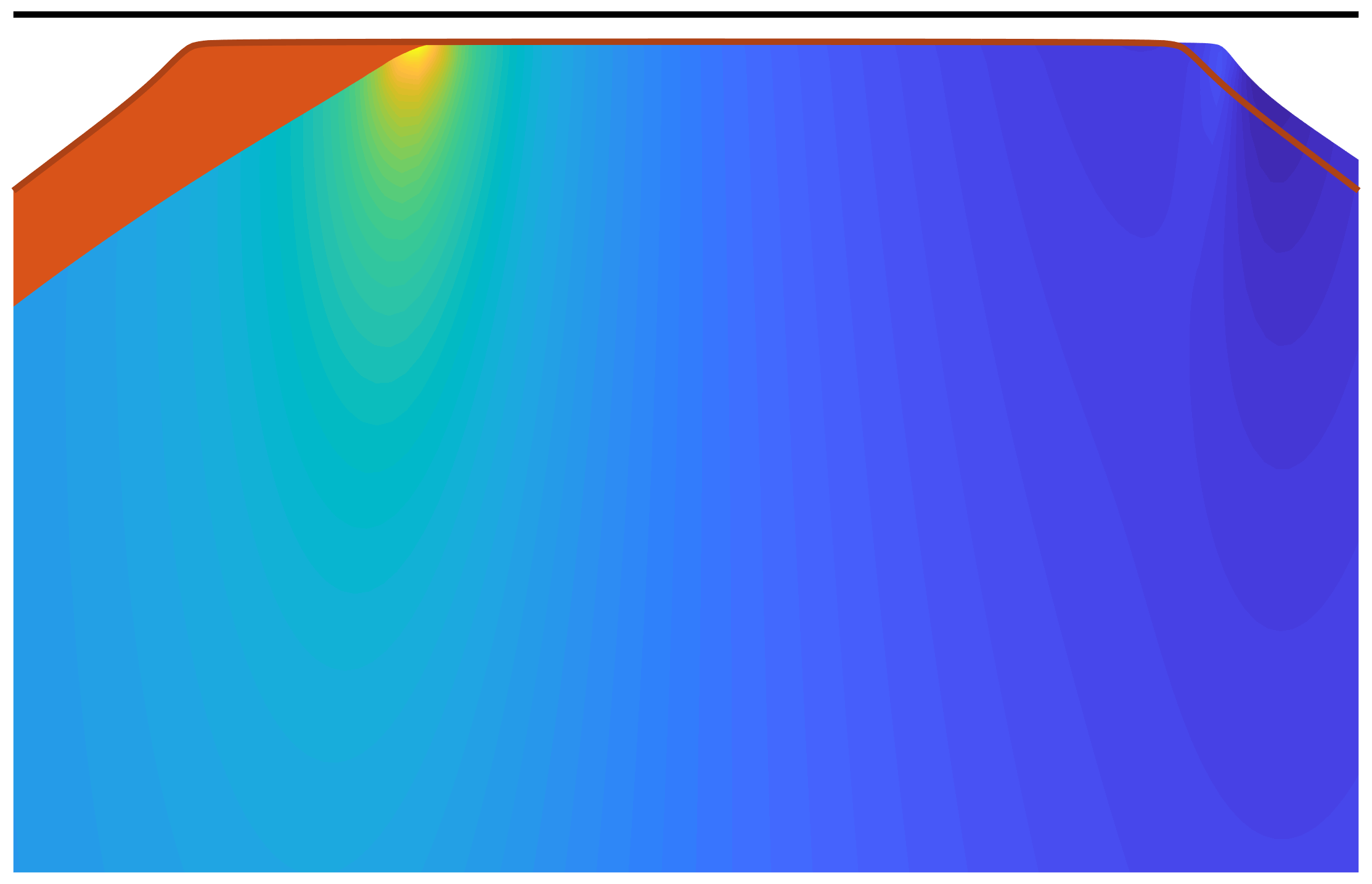}}
		\put(0.024,0.202){\small $\bar{u} = 0.5\,L_0$}
		\put(0.282,0.202){\small $\bar{u} = 1.0\,L_0$}
		\put(0.540,0.202){\small $\bar{u} = 1.5\,L_0$}
		\put(0.798,0.202){\small $\bar{u} = 2.0\,L_0$}
		\put(0.024,0.022){\small $\bar{u} = 2.5\,L_0$}
		\put(0.282,0.022){\small $\bar{u} = 3.0\,L_0$}
		\put(0.540,0.022){\small $\bar{u} = 3.5\,L_0$}
		\put(0.798,0.022){\small $\bar{u} = 4.0\,L_0$}
	\end{picture}
	\caption{Adhesive friction of a soft cap: Evolution of the contact
		interface during sliding for model~\CA\ and zero normal load (see the
		dots in \cref{f:cap:Ac0}); the vertical axis is stretched by the factor
		$\approx 3.93$; the configuration at zero tangential displacement
		($\bar{u} = 0$) as well as its contour are shown in orange; the other
		colors show the first invariant of the Cauchy stress in units of~$E$.}
	\label{f:cap:Aczoom}
\end{figure}
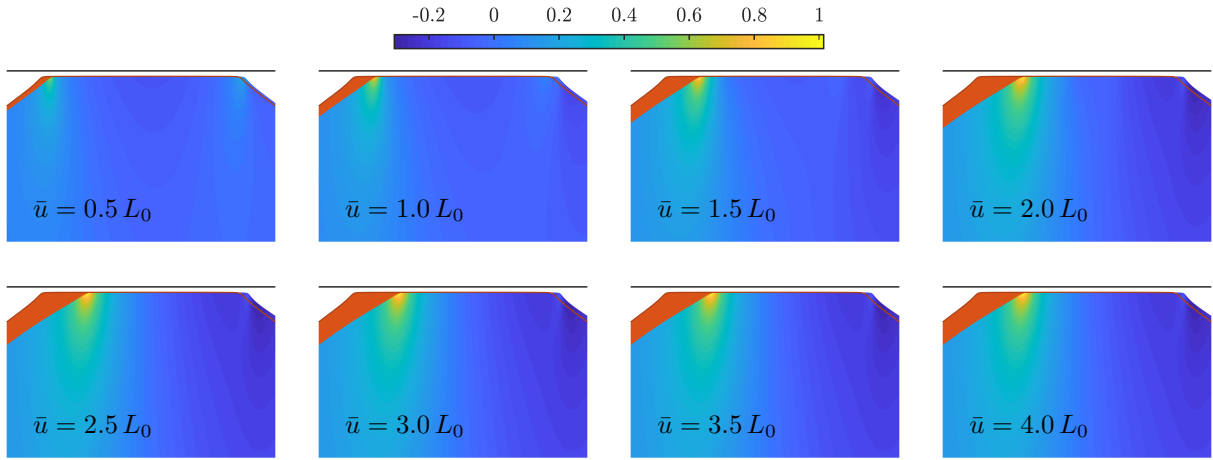
Finally, \cref{f:cap:Aczoom} shows (a vertically stretched view of) the contact
interface for model~\CA. As a comparison with the first four snapshots of
\cref{f:exp:snapshots} implies, model~\CA\ appropriately captures the
qualitative behavior of the contact edges before the friction force reaches its
static friction peak (see the squares in \cref{f:exp:results,f:cap:Ac0}).
Beforehand, mainly the left contact edge moves rightward with respect to the
base of the sphere. The subsequent leftward motion of the right edge (observed
in the experiments, see the last three snapshots in \cref{f:exp:snapshots}) is
not seen in the numerical results, which is also likely caused by the viscosity
of the material. In summary, combining model~\CA\ with a viscous material model
seems to be a promising approach to model sliding of smooth rubber spheres on
glass.


\section{Conclusion} \label{s:concl}

In this paper, we present two new continuum contact models for coupled adhesion
and friction that are suitable for soft systems like biological and
bio-inspired adhesives. These two models are based on a 3D continuum adhesion
formulation~\cite{sauer09cmame}, which is suitable to describe large
deformations and large sliding motions. As a motivation, we first provide a
review of existing experimental studies, theoretical investigations, and
modeling approaches for dry adhesion and friction. We also present new
experimental results for a smooth glass plate sliding on a smooth PDMS cap
under low normal loads. Our findings indicate that the classical law by Coulomb
and Amontons (with a linear dependence between the normal and tangential loads)
is not applicable for strong adhesion or small roughnesses. In this case the
friction force is also affected by an additional adhesion term. For some
applications, this additional term dominates, and the sliding resistance is
described by a constant, material-dependent frictional shear strength
multiplied with the real contact area. Our two continuum models can generate
sliding friction even when the normal pressure is zero or negative (i.e.,
tensile). This is useful because soft bio-adhesive pads are observed to
generate non-negligible sliding forces under zero normal load. The proposed
models contain the law of Coulomb and Amontons as a special case. As
demonstrated in~\cite{mergelPhD, mergel18comput}, for certain applications both
models show very similar behavior.

This paper focuses on the motivation and derivation of our new contact models.
Both their implementation in terms of a nonlinear finite element framework and
the algorithmic treatment of adhesive friction are discussed separately
\cite{mergelPhD, mergel18comput}. In~\cite{mergelPhD} the model is further
incorporated into a beam contact model~\cite{sauer14finel} for adhesive fibrils.

As an example we here investigate adhesive friction of a soft cap and a rigid
substrate, and show that the model behavior is in qualitative agreement to the
experiments. Our work is also a first important step towards a better
understanding and modeling of biologic adhesives: As shown in~\cite{mergelPhD},
model~\CA\ is also suitable to qualitatively describe the characteristics of
friction devices in stick insects. It would thus be very promising to perform
additional experiments on those, and to calibrate our friction models
accordingly.

In future work, several extensions should be developed to overcome the
restrictions mentioned in \cref{s:model:validity}, in particular the
restriction to small length scales that is inherent to numerical van der
Waals-based contact formulations. A contact homogenization technique should be
able to overcome this. The current comparison between model and experiment is
qualitative. With a refined model, it will be possible to conduct quantitative
comparisons, which will require further experiments with both biologic and
bio-inspired adhesives. It would also be interesting to apply model~\AM\ to
problems in which both adhesion- and pressure-controlled friction play a role.


\appendix

\section{Regularization for small normal distances} \label{a:regularize}

To avoid ill-conditioning, it is possible to regularize the contact stress for
small normal gaps, $\gn \to 0$, e.g.~by means of linear extrapolation. To this
end, we introduce $\Tn(\gn) \approx \Tn^\mathrm{reg}(\gn)$,
\begin{equation}
	\Tn^\mathrm{reg}(\gn) = \begin{cases}
		\Tn(\gn), & \gn \ge g_\mathrm{reg}, \\
		\Tn(g_\mathrm{reg}) + \Tn'(g_\mathrm{reg}) \, (\gn - g_\mathrm{reg}), &
		\gn < g_\mathrm{reg};
	\end{cases}
\end{equation}
see also \cref{f:Tn}. $\Tn'$ denotes the derivative of~$\Tn$ with respect
to~$\gn$. It is reasonable to choose a regularization distance of
$g_\mathrm{reg} \approx \gequ$, where $\gequ$ is the equilibrium distance. For
$g_\mathrm{reg} \le \gequ$, the attractive part of the curve in \cref{f:Tn}
remains unaffected; if $g_\mathrm{reg} > \gequ$, one also modifies the work of
adhesion, $W_\adh$, in \cref{e:Wadh}, and shifts~$\gequ$ to another position.


\section*{Acknowledgements}

We are grateful to the German Research Foundation (DFG) for supporting this
research under grants SA1822/5-1 and GSC111. We also thank Dr.~David Labonte
(University of Cambridge) for helpful comments, and Udit Pillai (RWTH Aachen
University) for his help with the FE mesh of the cylindrical cap. This work was
further supported by LABEX MANUTECH-SISE (ANR-10-LABX-0075) of Universit{\'e}
de Lyon, within the program Investissements d'Avenir (ANR-11-IDEX-0007)
operated by the French National Research Agency (ANR). In addition, it received
funding from the People Program (Marie Curie Actions) of the European Union's
Seventh Framework Program (FP7/2007--2013) under Research Executive Agency
Grant Agreement PCIG-GA-2011-303871. We are further indebted to Institut Carnot
Ing{\'e}nierie @ Lyon for support and funding.


{\footnotesize
\bibliographystyle{abbrv}
\bibliography{JournalsAbbr,Bibliography}
}

\end{document}